\documentclass[review,12pt]{elsarticle}

\usepackage{amssymb}
\usepackage{mathrsfs}
\usepackage{amsmath}
\usepackage{url}
\usepackage{float}
\usepackage{comment}
\usepackage{color}
\usepackage{multirow}
\usepackage{threeparttable}

\journal{}

\begin{document}

\begin{frontmatter}



\title{Mean Stress Tensor of Discrete Particle Systems in Submerged Conditions}

\author[First,Second]{Zhuan Ge}

\author[Second]{Teng Man\corref{cor1}}
\ead{manteng@westlake.edu.cn}
\cortext[cor1]{Corresponding author.}
\author[Second]{Sergio Andres Galindo-Torres\corref{cor1}}
\ead{s.torres@westlake.edu.cn}


\address[First]{ Zhejiang University, Hangzhou, Zhejiang 310058, China}
\address[Second]{School of Engineering, Westlake University, 18 Shilongshan Street, Hangzhou, Zhejiang 310024, China}





\begin{abstract}
The mean stress tensor is essential to investigate the dynamics of granular material. In this paper, we use Hamilton's principle of least action to derive the averaged stress tensor of discrete granular assemblies subjected to hydraulic force fields, as well as rigorous conditions for a proper definition of the Representative Volume Element (RVE). The main goal behind our efforts is to upscale particle physics into a sound stress tensor for systems involving the complex interaction between grains and water. We identify the contributions from the unbalanced forces, hydraulic forces, gravity, external forces, and particle fluctuation to the mean stress tensor. In doing so, it is convenient to separate the influence of different force fields when the granular system is subjected to complex environments, e.g., subaqueous conditions. The obtained formula is then validated by triaxial test simulations of dry and saturated granular systems using the Discrete Element Method (DEM) and the Lattice-Boltzmann Method (LBM). The results show that the deduced formula can accurately calculate the stress tensor of discrete assemblies with various body-force fields. We used validated DEM-LBM simulations of submerged granular column collapses to explore the physics happening at the grain scale with this mathematical formalism and showcase its potential. We provide a new perspective based on the granular assembly scale to pursue the fluid-solid interaction. Due to the importance of stress analysis in the constitutive modelling of granular materials, this work could help to better obtain the stress-strain relationship of saturated or submerged granular systems.
\end{abstract}

\begin{graphicalabstract}
\includegraphics[scale=0.32]{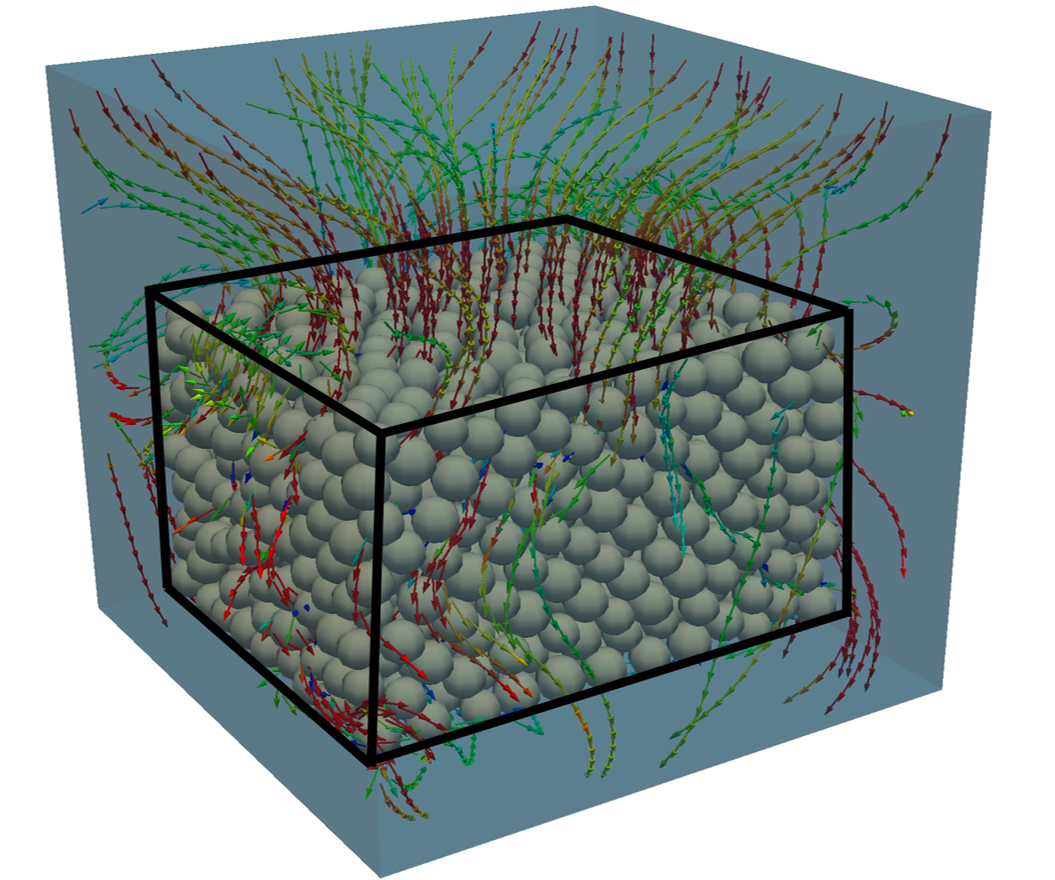}
\end{graphicalabstract}

\begin{highlights}
\item Derivation and validation of the averaged stress for submerged granular condition.
\item Quantification of the contribution of various force fields to the averaged stress.
\item Verification of the difference between pore water pressure and the hydrodynamic contribution to the effective stress in the submerged granular assembly. 
\end{highlights}


\end{frontmatter}


\section{Introduction} \label{section:Intro}
Granular materials are ubiquitous in natural and engineering systems, such as fresh concrete, debris flows, landslides, and particulate flows in chemical engineering and food processing  \cite{sousa2019drying,cao2017numerical,seguin2011dense,vo2020additive,miller2017tsunamis}. The stress analysis is vital for understanding the nature of granular systems under various loading conditions \cite{shi2020homogenization,vo2020additive,laubie2017stress}. There are currently two different approaches to modelling the behavior of granular materials: (i) the Discrete Element Method(DEM), where each grain is described explicitly to represent the microscopic behavior of the system; (ii) continuum approaches, where granular assemblies are treated as continuum materials and then solved using momentum balance equations with various constitutive laws (e.g., Navier-Stokes equations with various non-Newtonian fluid assumptions, Mohr-Coulomb theory in soil mechanics, etc.).

It is convenient to implement continuum models to investigate the macroscopic behavior of granular materials such as in references \cite{franci20193d,baumgarten2019general}, especially for disasters such as submarine avalanches or debris flows \cite{hsu2010applications,salvatici2017debris}. Due to the large scale and fast transport speed, it is challenging to study these natural hazards at the particle scale directly. Numerous researchers \cite{liu2006numerical,uchida2021role,nishiguchi2011prediction} used the continuum models to investigate these disasters, but choosing the correct stress-strain relationship is the key for the continuum model to simulate the granular materials in various conditions. Since granular assemblies can flow like fluid or solid, the stress-strain relationship becomes more complicated.  When the granular assembly flow like a fluid, Jop et at.\cite{jop2006constitutive} proposed a constitutive law based on the rheology of dry granular assembly to describe the stress-strain relationship. Baumgarten et at.\cite{baumgarten2019general} presented a constitutive model for fluid-saturated sediments transport using the viscous inertial rheology of submerged wet grains.
The stress-strain relationship of granular assemblies could be verified from DEM simulations. Based on the advantage of the DEM simulation, Guo et al.\cite{guo2014coupled} proposed the DEM coupled with the Finite Element Method(FEM) to investigate the behavior of granular media using a multi-scale approach, where no constitutive equation is assumed, and the stress-strain relation is obtained directly from DEM. It is the authors' opinion that this strategy is sound, however, it is challenging to obtain the accurate stress field from the discrete granular system once complex force fields, such as hydrodynamic force fields or electromagnetic forces, are present. This work pretends to close that knowledge gap.

\begin{figure}
	\centering
	\includegraphics[scale=0.32]{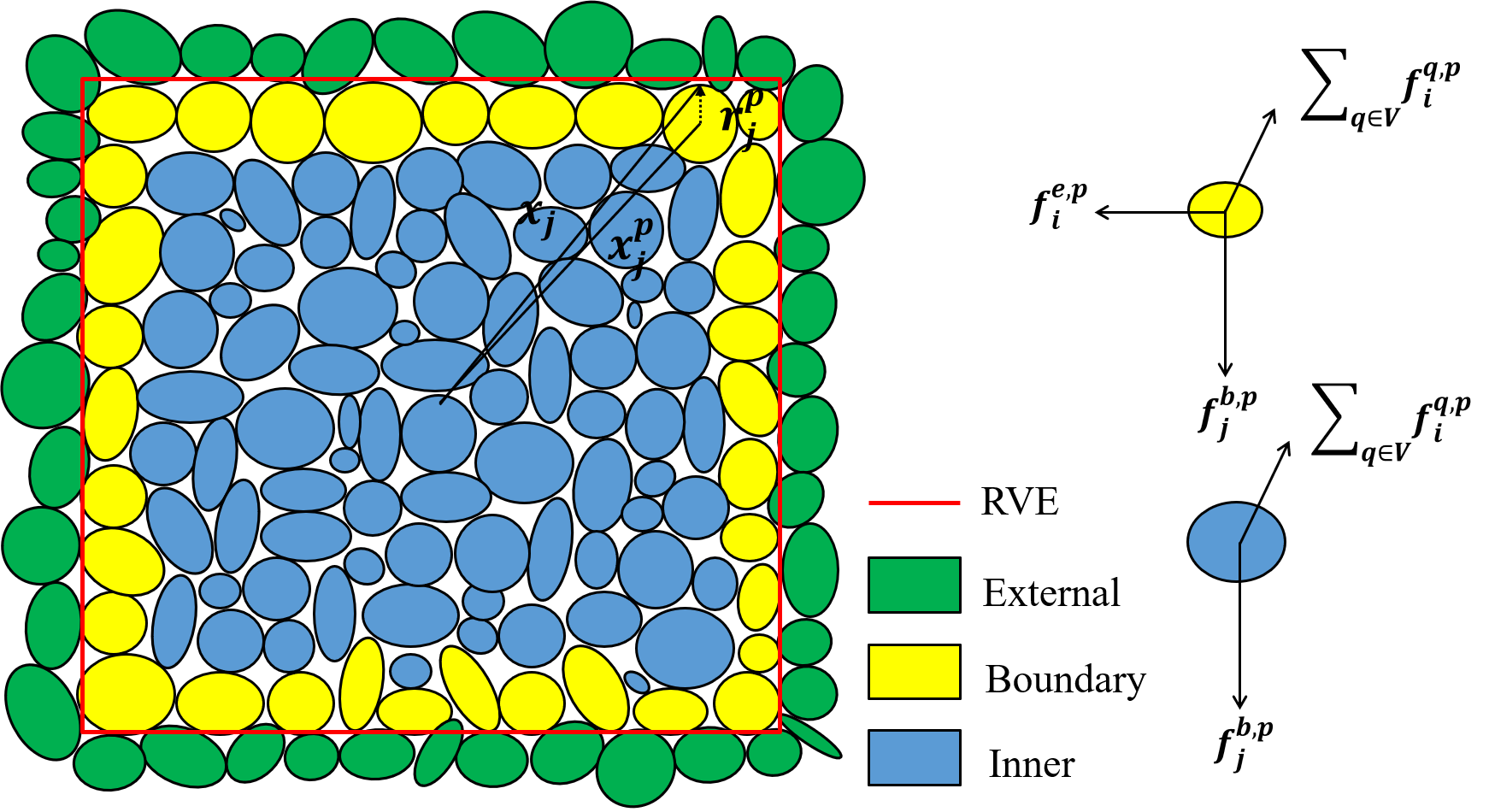}
	\caption{Force analysis of the particle both inside the domain and on the boundary}
	\label{Forceanalysis}
\end{figure}

The problem of accurately obtaining the correct stress-strain relation from DEM simulations in the granular flow system requires defining an appropriate volume. At first, the granular assembly can be divided into several representative volume elements (RVE), usually in a cubic grid; secondly, the macroscopic information such as stress and strain of each RVE is obtained using the microscopic particle-particle information; finally, the stress-strain relation of the whole granular system can be investigated at a different time and different location.
As shown in Fig. \ref{Forceanalysis}, an RVE, $V$ consists of $N$ discrete particles with different shapes and material types. The particles can be divided into two groups: the boundary part ($p \in \partial V$) and the internal part ($p \in V-\partial V$). The boundary particles are subjected to the body forces $f_{i}^{b,p}$ (such as gravity, hydrodynamic, or electromagnetic forces in different conditions), external forces $f_{i}^{e,p}$, and contact forces $\sum_{q \in V} f_{i}^{q,p}$ from the surrounding particles ($q\in V$) inside the volume($i=x,y,z$ to components of the vector), the inner particles are only subjected to body forces and contact forces. 
In previous works \cite{weber1966recherches,christoffersen1981micromechanical,bagi1996stress,bardet2001asymmetry,goldhirsch2002microscopic,bagi1999microstructural,chang2005virtual}, the expression of the averaged stress tensor of the granular assembly is based on the contact forces $f_{i}^{c}$(equal to $\sum_{q \in V} f_{i}^{q,p}$) and branch vectors $l_{j}^{c}$(equal to the vector going from particle $p$'s center of mass to particle $q$'s) under the assumption of quasi-static equilibrium conditions. This formula is widely used in many fields to obtain the averaged stress tensor,
\begin{equation}
\langle \sigma_{ij} \rangle_{\text{Weber}}=\frac{1}{V}\sum_{c=1}^{N_{c}}f_{i}^{c} l_{j}^{c}, \qquad i,j = x,y,z \label{equa:previous}
\end{equation}
where $N_{c}$ is the number of contact pairs inside the sampling volume.

Bagi et al.\cite{bagi1999microstructural} investigated the influence of gravity on the average stress and proposed that gravity was already implicitly included in the contact force pair at equilibrium. Fortin et al.\cite{fortin2003construction,de2004numerical} proposed an averaged stress tensor formula for the granular medium, which includes the effect of body forces. This formula includes the body forces term explicitly, this is different from the work of Bagi\cite{bagi1999microstructural}. They pointed out that body forces and inertial effects are essential for the Cauchy theory. Nicot et al.\cite{nicot2013definition} studied the influence of the inertial term on the averaged stress tensor. They gave the expression of the averaged stress tensor including two terms: The first term is in agreement with the work of Weber\cite{bagi1999microstructural}, which includes the contribution of body force implicitly. The second term consists of the dynamic effects related to the rotations and accelerations of the particles. 

Yan et al.\cite{yan2019definition} investigated the importance of the boundary-radius-gap term and suggested that the boundary-radius-gap term should be considered in the averaged stress tensor. The boundary-radius-gap term is defined as $\frac{1}{V}\sum_{p \in \partial V}^{}f_{i}^{e,p}r_{j}^{p}$, where $r_{j}^{p}$ is the boundary radius gap vector from the particle center of mass to the contact point with the boundary, as shown in Fig. \ref{Forceanalysis}. Since the external force and the boundary radius gap vector are unknown in the RVE, previous works such as in \cite{weber1966recherches,bagi1996stress,fortin2003construction,nicot2013definition} need to use a large enough number of particles to ignore the effect of this term, which affects the robustness of the implementation of RVE. So it is necessary to propose a method to obtain the boundary-radius-gap term using the internal granular contact information. 

Goldhirsch et al.\cite{goldhirsch2002microscopic} proposed the standard coarse graining theory based on the kinetic theory to calculate the stress field of the granular materials, however, this method treats the granular as a point with mass, which means no granular volume and granular shape effects. The coarse graining theory is not suitable for the macro-scale granular system in which particle has different shapes and the volume can not be neglected (such as the presence of boundary-radius-gap term).

As shown above, the definition of the averaged stress tensor over a granular assembly is still a debating topic,
especially for the granular assembly subjected to different force fields, such as hydrodynamic forces for submerged granular media, capillary forces for unsaturated soils, and gravitational force field for gravity-driven currents, hence a universal definition of the averaged stress tensor is necessary.

This paper uses Hamilton's principle to derive the averaged stress tensor for discrete assemblies subjected to different force fields (including gravity and hydrodynamic forces) and presents a method to obtain the boundary-radius-gap term which is neglected by the previous studies\cite{christoffersen1981micromechanical,bagi1996stress,bagi1999microstructural,nicot2013definition,fortin2003construction} using microscopic information inside the RVE. The contribution of each force field to the averaged stress tensor can be separated into different terms. In this way, we can investigate the effect of a specific force field on the assembly. First, we review how previous work calculates the averaged stress tensor of granular assembly subjected to different force fields. Then, in Section.\ref{Section:DES}, an analytical derivation is implemented to define the averaged stress tensor of the granular assemblies subjected to various body force fields. A method is proposed to obtain the boundary-radius-gap term. In Section.\ref{sectionnum}, dry and saturated compression tests are implemented to examine the proposed formula using DEM and DEM coupled with Lattice Boltzmann Method (LBM-DEM) simulations. The deduced averaged stress are applied to investigate the influence of fluid on the granular during the transient submerged granular column collapse and get insight into the physics happening during this process. Finally, Section.\ref{sec:conclu} summarized this work and proposed potential avenues to use this contribution for the formulation of constitutive models of complex granular flows.

\section{Derivations of averaged stress} \label{Section:DES}

We use Hamilton's principle of least action to derive the expression of the stress tensor of granular media subjected to various force fields, as shown in Section.\ref{section:Intro}. The lagrangian of an elastic system can be expressed as
\begin{equation}
L=E_{k}-(V_{\varepsilon}-W),
\end{equation}
where $E_k$ is the kinetic energy, $V_{\varepsilon}$ is the strain energy, and $W$ is the external work. According to Hamilton's principle of least action($\Gamma$ being the action),
\begin{equation}
\delta \Gamma=\int_{t_{1}}^{t_{2}} \delta L dt=\int_{t_{1}}^{t_{2}}\delta[E_{k}-(V_{\varepsilon}-W)]dt=0,
\end{equation}
where time $t$ belongs to a infinitesimal interval $[t_{1},t_{2}]$. The external virtual work consists of both the body force work $W_{b}$ and the external surface force work $W_{e}$ as follows
\begin{equation}
\delta W=\delta W_{b}+\delta W_{e}.
\end{equation}
We can obtain the virtual work balance equation,
\begin{equation}
\int_{t_{1}}^{t_{2}}\delta E_{k}-(\delta V_{\varepsilon}-\delta W_{b}-\delta W_{e})dt=0. \label{equa:continvir}
\end{equation}
The macroscopic granular system is treated as a Cosserat continuum, each point of the media can be represented as an infinitesimal rigid body\cite{ioannis2018cosserat}. Hence, both displacements and rotations exist. The virtual displacements and rotations $\delta u_{i}$ and $\delta \theta_{i}$ can be selected arbitrarily. In particular, they can be given by \cite{bardet2001asymmetry}
\begin{subequations}
\begin{align}
&\delta u_{i}= a_{i}+b_{ij}x_{j}+c_{ijk}x_{j}x_{k}, \qquad   i,j,k=x, y, z,\label{equa:stressu}\\
&\delta \theta_{i}=\alpha_{i}+\beta_{ij}x_{j},  \qquad  \qquad \qquad \; \,  i,j,k=x, y, z.\label{equa:stresst}
\end{align}
\end{subequations}
where $a_{i}$, $b_{ij}$, $c_{ijk}$, $\alpha_{i}$, and $\beta_{ij}$ are arbitrary coefficients. The variational form of the kinetic energy in Eq.\ref{equa:continvir} is then calculated as
\begin{equation}
   \int_{t_{1}}^{t_{2}}\delta E_{k}dt=\int_{t_{1}}^{t_{2}}\int_{V}\rho \dot{u_{i}}\delta \dot{u_{i}}+\rho \kappa_{ij}\dot{\theta}_{i}\delta\dot{\theta}_{i}dVdt
\end{equation}
where $\dot{u}_{i}$, $\dot{\theta}_{i}$, $\rho$, and $\rho \kappa_{ij}\dot{\theta}_{j}$ are the translational velocity, internal rotational velocity, density, and the internal spin density of the material point. $\kappa_{ij}$ is the micro-inertia tensor as is defined in \cite{ioannis2018cosserat}. Using the integration by parts and noting that the virtual displacements and rotations at initial and final times are nil, $\delta \dot{u}_{i}(t_{1})=\delta \dot{u}_{i}(t_{2})=0$, $\delta \dot{\theta}_{i}(t_{1})=\delta \dot{\theta}_{i}(t_{t_{2}})=0$), hence, one can obtain
\begin{equation}
   \int_{t_{1}}^{t_{2}}\delta E_{k}dt=-\int_{t_{1}}^{t_{2}}\int_{V}\rho \Ddot{u}_{i}\delta u_{i} +\rho \kappa_{ij}\Ddot{\theta}_{j}\delta \theta_{i} dVdt.
\end{equation}
According to Reynold's transport theorem, the material acceleration is given by the local acceleration and the convective acceleration\cite{tadmor2011modeling}, hence, the variational kinetic energy is 
\begin{equation}
    \delta E_{k}=-\int_{V}\frac{\partial \rho \dot{u}_{i}}{\partial t}\delta u_{i}+(\rho \dot{u}_{i}\dot{u}_{j})_{,j}\delta u_{i} +\frac{\partial \rho \kappa_{ij}\dot{\theta}_{j}}{\partial t}\delta \theta_{i}+(\rho \dot{u}_{i}\kappa_{ij}\dot{\theta})_{,j}\delta \theta_{i} dV,\\
\end{equation}
Then, taking advantage of the Gauss theorem for $\int_{V}(\rho \dot{u}_{i}\dot{u}_{j})_{,j}\delta u_{i}+(\rho \dot{u}_{i}\kappa_{ij}\dot{\theta})_{,j}\delta \theta_{i}dV$, and with the assumption of homogeneous micro-deformations for the material, the surface integrals are nil, we can obtain
\begin{equation}
    \delta E_{k}=-\int_{V}\rho \Ddot{u}_{i}\delta u_{i}-(\rho \dot{u}_{i}\dot{u}_{j})\delta u_{i,j} +\rho \kappa_{ij}\Ddot{\theta}_{j}\delta \theta_{i}-(\rho \dot{u}_{i}\kappa_{ij}\dot{\theta}_{i})\delta \theta_{i,j} dV.
\end{equation}
In the case of the discrete granular assembly, the virtual kinetic energy is given by
\begin{equation}
\begin{split}
    \delta E_{k}
    &=-\sum_{p \in V}\int_{V_{p}}\rho \Ddot{u}_{i}\delta u_{i}-(\rho \dot{u}_{i}\dot{u}_{j})\delta u_{i,j} +\rho \kappa_{ij}\Ddot{\theta}_{j}\delta \theta_{i}-(\rho \dot{u}_{i}\kappa_{ij}\dot{\theta}_{i})\delta \theta_{i,j} dV_{p}\\
    &=-\sum_{p \in V}[f_{i}^{t,p}(a_{i}+b_{ij}x_{j}^{p}+c_{ijk}x_{j}^{p}x_{k}^{p})-m^{p}v_{i}^{'}v_{j}^{'}(b_{ij}+c_{ijk}x_{k}^{p})\\
    &+\mathcal{M}_{i}^{t,p}(\alpha_{i}+\beta_{ij})-v_{i}^{'}I_{ij}\dot{\theta}_{j}(\beta_{ij})]
\end{split}\label{equa:EK}
\end{equation}
where $x_{j}^{p}=x_{j}^{V_{c}}-x_{j}^{G_{p}}$ is the vector from mass center $x_{j}^{G_{p}}$ of the particle to the  center $x_{j}^{V_{c}}$ of the RVE, $f_{i}^{t,p}$, $v_{i}^{'}=v_{i}^{p}-v_{i}^{V_{cm}}$, $\dot{\theta}_{j}$, and $I_{ij}$ are the unbalance force, fluctuation velocity, spin velocity, and moment of inertia of the particle. $v_{i}^{p}$ is the translational velocity of particle, $v_{i}^{V_{cm}}=\frac{1}{N}\sum_{V}v_{i}^{p}$ is the averaged velocity of the RVE. $\mathcal{M}_{i}^{t,p}$ is the resultant moment of the particle.
The virtual work of the external force in Eq.\ref{equa:continvir} from a continuum system to a discrete system is
\begin{equation}
\begin{split}
\delta W_{e}&=\int_{S} f_{i}^{e}\delta u_{i}+\mathcal{M}_{i}^{e}\delta \theta_{i}dS\\&=\int_{S} f_{i}^{e}(a_{i}+b_{ij}x_{j}+c_{ijk}x_{j}x_{k})+\mathcal{M}_{i}^{e}(\alpha_{i}+\beta_{ij}x_{j})dS\\
&=\sum_{p \in S}f_{i}^{e,p}[a_{i}+b_{ij}(x_{j}^{p}+r_{j}^{p})+c_{ijk}(x_{j}^{p}+r_{j}^{p})(x_{k}^{p}+r_{k}^{p})]\\
&+\mathcal{M}_{i}^{e,p}[\alpha_{i}+\beta_{ij}x_{j}^{p}],\\
\end{split}\label{equ:we}
\end{equation}
where $f_{i}^{e}$, and $\mathcal{M}_{i}^{e}$ are the forces on the boundary surface point and the moment exerted by the boundary force. $f_{i}^{e,p}$ and $\mathcal{M}_{i}^{e,p}$ are the external force exerted on the particle and the moments exerted by external contact of the particle, $r_{j}^{p}=x_{j}^{e}-x_{j}^{G_{p}}$ is the boundary radius gap vector from the mass center of particle to the external contact point $x_{j}^{e}$, and $S$ represents the boundary surface. The virtual work of the body force in Eq.\ref{equa:continvir} is written as
\begin{equation}
\begin{split}
\delta W_{b}&=\int_{V} f_{i}^{b}\delta u_{i}+\mathcal{M}_{i}^{b}\delta \theta_{i}dV\\&=\int_{V} f_{i}^{b} (a_{i}+b_{ij}x_{j}+c_{ijk}x_{j}x_{k})+\mathcal{M}_{i}^{b}(\alpha_{i}+\beta_{ij}x_{j})dV\\
&=\sum_{p \in V}f_{i}^{b,p}(a_{i}+b_{ij}x_{j}^{p}+c_{ijk}x_{j}^{p}x_{k}^{p})+\mathcal{M}_{i}^{b,p}[\alpha_{i}+\beta_{ij}(x_{j}^{p})], \label{equa:wb}
\end{split}
\end{equation}
where $f_{i}^{b}$, $\mathcal{M}_{i}^{b}$ are the body force and internal moments due to $f_{i}^{b}$ of the material point, $f_{i}^{b,p}$, and $\mathcal{M}_{i}^{b,p}$ are the body force and moment for each particle. The body force can vary in different conditions, such as hydrodynamic forces, gravity, and electromagnetic forces. For convenient, we use $f_{i}^{b,p}$ and $\mathcal{M}_{ij}$ to give a general derivation. In a Cosserat continuum\cite{bardet2001asymmetry,chang2005virtual}, the strain energy can be expressed as
\begin{equation}
\begin{split}
    \delta V_{\varepsilon}&=\int_{V} \sigma_{ij} (\delta u_{i,j}+e_{ijk}\delta \theta_{k})+\mu_{ij}\delta \theta_{i,j} dV \\
    &=\int_{V} \sigma_{ij}(b_{ij}+c_{ijk}x_{k})+\sigma_{ij}e_{ijk}(\alpha_{i}+\beta_{ij}x_{j})+\mu_{ij}\beta_{ij}dV\\
    &=b_{ij}\int_{V}\sigma_{ij}dV+c_{ijk}\int_{V}(\sigma_{ij}x_{k}+\sigma_{ki}x_{j})dV-\alpha_{i} e_{ijk}\int_{V}\sigma_{jk}dV\\
    &+\beta_{ij}\int_{V}(\mu_{ji}+e_{ikl}x_{j})dV
\end{split}\label{equ:wve}
\end{equation}
where $\sigma_{ij}$ is the stress tensor that is related to the linear displacement of the material, $\mu_{ij}$ is the couple stress tensor referring to the internal rotation, $e_{ijk}$ is the permutation symbol used for vector cross-product. According to Eq.\ref{equa:continvir}, we got the following relationship 
\begin{equation}
\delta V_{\varepsilon}=\delta E_{k}+\delta W_{b}+\delta W_{e}.
\end{equation}
Combining with Eqs.\ref{equa:EK}, \ref{equ:we}, \ref{equa:wb}, and \ref{equ:wve}, and noting that
Eqs.\ref{equa:stressu} and.\ref{equa:stresst} hold for arbitrary values of $a_{i}$, $b_{ij}$, $c_{ijk}$, $\alpha_{i}$, and $\beta_{ij}$, the following relations are presented
\begin{subequations}
	\begin{align}
	&0=a_{i}\sum_{p \in V}(-f_{i}^{t,p}+f_{i}^{b,p}+f_{i}^{e,p}),\label{equa:plm}\\
	&b_{ij}\int_{V} \sigma_{ij}dV=b_{ij}\sum_{p \in V}(-f_{i}^{t,p}x_{j}^{p}+f_{i}^{b,p}x_{j}^{p}+f_{i}^{e,p}x_{j}^{p}+f_{i}^{e,p}r_{j}^{p}-m^{p} v_{i}^{'}v_{j}^{'}),\label{equa:tensor}\\ 
	&c_{ijk}\int_{V}(\sigma_{ij}x_{k}+\sigma_{ki}x_{j})dV
 =c_{ijk}\sum_{p \in V}[-f_{i}^{t,p}x_{j}^{p}x_{k}^{p}+f_{i}^{b,p}x_{j}^{p}x_{k}^{p}+f_{i}^{e,p}(x_{j}^{p}+r_{j}^{p})(x_{k}^{p}+r_{k}^{p})+m^{p}v_{i}^{'}v_{j}^{'}x_{k}^{p}],\label{equa:ht}\\
    &\alpha_{i}e_{ijk}\int_{V}\sigma_{jk}dV=\alpha_{i}\sum_{p \in V}(-\mathcal{M}_{i}^{t,p}+\mathcal{M}_{i}^{b,p}+\mathcal{M}_{i}^{e,p}),\label{equa:SM}\\
    &\beta_{ij}\int_{V}(\mu_{ij}+e_{jkl}x_{i}\sigma_{lk})dV=\beta_{ij}\sum_{p \in V}(-\mathcal{M}_{i}^{t,p}x_{j}^{p}+\mathcal{M}_{i}^{b,p}x_{j}^{p}+\mathcal{M}_{i}^{e,p}x_{j}^{p}-v_{i}^{'}I_{ij}\dot{\theta}_{j}),\label{equa:Ctensor}
	\end{align} 
\end{subequations}
Eq.\ref{equa:plm} shows the sum of the force balance from a particle scale, Eq.\ref{equa:ht} gives the high order stress $\langle \sigma_{ijk}\rangle=\frac{1}{2V}\int_{V}\sigma_{ij}x_{k}+\sigma_{ki}x_{j}dV$ as is defined in Ref.\cite{chang2005virtual}, volume averaging of Eq.\ref{equa:SM} is the averaged macro internal moment $\langle \mathcal{M}_{i}^{T} \rangle=\frac{1}{V}e_{ijk}\int_{V}\sigma_{jk}dV$, the expression of the averaged stress tensor, $\langle\sigma_{ij}\rangle$, and averaged couple stress, $\langle\mu_{ij}\rangle$, are given from Eqs.\ref{equa:tensor} and \ref{equa:Ctensor},
\begin{subequations}
\begin{align}
        &\langle \sigma_{ij}\rangle=\frac{1}{V}\int_{V} \sigma_{ij}dV
        =\frac{1}{V}\sum_{p \in V}(-f_{i}^{t,p}x_{j}^{p}+f_{i}^{b,p}x_{j}^{p}+f_{i}^{e,p}x_{j}^{p}+f_{i}^{e,p}r_{j}^{p}-m^{p}v_{i}^{'}v_{j}^{'})\label{equ:divide}.\\
    &\langle \mu_{ij}\rangle=\frac{1}{V}\int_{V} \mu_{ij}+e_{jkl}x_{i}\sigma_{lk}dV=\frac{1}{V}\sum_{p \in V}(-\mathcal{M}_{i}^{t,p}x_{j}^{p}+\mathcal{M}_{i}^{b,p}x_{j}^{p}+\mathcal{M}_{i}^{e,p}x_{j}^{p}-v_{i}^{'}I_{ij}\dot{\theta}_{j}),
\end{align}
\end{subequations} 
It is worth noting that $f_{i}^{e,p}=0$ $(p\in V-\partial V)$ as shown in Fig. \ref{Forceanalysis}, the forces and moments subjected to a particle $p \in V$ can be expressed as
\begin{subequations}
\begin{align}
    &\sum_{q \in V} f_{i}^{q,p}+f_{i}^{e,p}+f_{i}^{b,p}=\frac{1}{V}\sum_{p \in V}f_{i}^{t,p} \label{froceana}\\
    &\sum_{q \in V} \mathcal{M}_{i}^{q,p}+\mathcal{M}_{i}^{e,p}+\mathcal{M}_{i}^{b,p}=\frac{1}{V}\sum_{p \in V}\mathcal{M}_{i}^{t,p} \label{froceanM}
\end{align}
\end{subequations}
where $f_{i}^{q,p}$ and $\mathcal{M}_{i}^{q,p}$ are the contact force and moment subjected by particle $p$ from particle $q$. Combining Eq.\ref{equ:divide} and \ref{froceana}, the averaged stress tensor can be written as
\begin{subequations}
    \begin{align}
        &\langle \sigma_{ij}\rangle=-\frac{1}{V}\sum_{p \in V} \sum_{q \in V} f_{i}^{q,p}x_{j}^{p}+\frac{1}{V}\sum_{p \in V}f_{i}^{e,p}r_{j}^{p}-\frac{1}{V}\sum_{p\in V}m^{p}v_{i}^{'}v_{j}^{'}.\\
        &\langle \mu_{ij}\rangle=-\frac{1}{V}\sum_{p \in V} \sum_{q \in V} \mathcal{M}_{i}^{q,p}x_{j}^{p}-v_{i}^{'}I_{ij}\dot{\theta}_{j}.
    \end{align}
\end{subequations}
Noting that, $\sum_{p \in V}\sum_{q \in V} f_{i}^{q,p}(x_{j}^{V_{c}}-x_{j}^{G_{p}})=\sum_{q \in V}\sum_{p \in V} f_{i}^{p,q}(x_{j}^{V_{c}}-x_{j}^{G_{q}})$, $\sum_{p \in V}\sum_{q \in V} \mathcal{M}_{i}^{q,p}(x_{j}^{V_{c}}-x_{j}^{G_{p}})=\sum_{q \in V}\sum_{p \in V} \mathcal{M}_{i}^{p,q}(x_{j}^{V_{c}}-x_{j}^{G_{q}})$and $f^{q,p}=-f^{p,q}$, $\mathcal{M}^{q,p}=-\mathcal{M}^{p,q}$, hence, one can obtain
\begin{equation}
\begin{split}
       -\sum_{p \in V}\sum_{q \in V} f_{i}^{q,p}(x_{j}^{V_{c}}-x_{j}^{G_{p}})&=\sum_{p \in V}\sum_{q \in V} f_{i}^{q,p}(x_{j}^{V_{c}}-x_{j}^{G_{q}})\\&=\frac{1}{2}\sum_{p \in V}\sum_{q \in V} f_{i}^{q,p}(x_{j}^{G_{q}}-x_{j}^{G_{p}})
\end{split}
\end{equation}
\begin{equation}
\begin{split}
    -\sum_{p \in V}\sum_{q \in V} \mathcal{M}_{i}^{q,p}(x_{j}^{V_{c}}-x_{j}^{G_{p}})&=\sum_{p \in V}\sum_{q \in V} \mathcal{M}_{i}^{q,p}(x_{j}^{V_{c}}-x_{j}^{G_{q}})\\&=\frac{1}{2}\sum_{p \in V}\sum_{q \in V} \mathcal{M}_{i}^{q,p}(x_{j}^{G_{q}}-x_{j}^{G_{p}})
\end{split}
\end{equation}
where $f_{i}^{c}=f_{i}^{q,p}$, $\mathcal{M}_{i}^{c}=\mathcal{M}_{i}^{q,p}$ and $l_{j}^{c}=x_{j}^{G_{q}}-x_{j}^{G_{p}}$, thus the contribution of contact force can also be expressed as
\begin{subequations}
\begin{align}
    &-\frac{1}{V}\sum_{q \in V} f_{i}^{q,p}x_{j}^{p}=\frac{1}{V}\sum_{c=1}^{N_{c}} f_{i}^{c}l_{j}^{c}.\label{equa:lc}\\
    &\-\frac{1}{V}\sum_{q \in V} \mathcal{M}_{i}^{q,p}x_{j}^{p}=\frac{1}{V}\sum_{c=1}^{N_{c}} \mathcal{M}_{i}^{c}l_{j}^{c}.
\end{align}
\end{subequations}
\begin{table*}
\begin{center}
\caption{\label{tab:table1} Contributions of each force field to the averaged stress and averaged couple stress.}
\begin{tabular}{ccc}
\hline
\multicolumn{1}{l}{\multirow{2}{*}{Averaged stress}}&\multicolumn{2}{l}{$\langle \sigma_{ij}\rangle=\sigma_{ij}^{t}+\sigma_{ij}^{b}+\sigma_{ij}^{e}+\sigma_{ij}^{r}+\sigma_{ij}^{k}$} \\
&$\quad \quad \;=\sigma_{ij}^{c}+\sigma_{ij}^{r}+\sigma_{ij}^{k}$\\
\multicolumn{1}{l}{\multirow{2}{*}{Averaged couple stress}}&\multicolumn{2}{l}{$\langle \mu_{ij} \rangle=\mu_{ij}^{t}+\mu_{ij}^{b}+\mu_{ij}^{e}+\mu_{ij}^{k}$}\\
&$=\mu_{ij}^{c}+\mu_{ij}^{k}$\\
\hline
Description & Translational & Rotational\\
\hline
\multicolumn{3}{c}{Applied}\\
External surface force  &$\sigma_{ij}^{e}=\frac{1}{V}\sum_{p\in V}f_{i}^{e,p}x_{j}^{p}$& $\mu_{ij}^{e}=\frac{1}{V}\sum_{p\in V}\mathcal{M}_{i}^{e,p}x_{j}^{p}$\\
 Boundary radius gap&$\sigma_{ij}^{r}=\frac{1}{V}\sum_{p\in V}f_{i}^{e,p}r_{j}^{p}$& - \\
 Body force (eg: hydraulic)&$\sigma_{ij}^{b}=\frac{1}{V}\sum_{p\in V}f_{i}^{e,p}x_{j}^{p}$& $\mu_{ij}^{b}=\frac{1}{V}\sum_{p\in V}\mathcal{M}_{i}^{b,p}x_{j}^{p}$\\
 \hline
 \multicolumn{3}{c}{Reaction}
 \\
 Contact force&$\sigma_{ij}^{c}=\frac{1}{V}\sum_{p\in V}f_{i}^{c}l_{j}^{c}$& $\mu_{ij}^{c}=\frac{1}{V}\sum_{p\in V}\mathcal{M}_{i}^{c}l_{j}^{c}$\\
Unbalance force&$\sigma_{ij}^{t}=-\frac{1}{V}\sum_{p\in V}f_{i}^{t,p}x_{j}^{p}$& $\mu_{ij}^{t}=-\frac{1}{V}\sum_{p\in V}\mathcal{M}_{i}^{t,p}x_{j}^{p}$\\
Kinetic fluctuation&$\sigma_{ij}^{k}=-\frac{1}{V}\sum_{p\in V}m^{p}v_{i}^{'}v_{j}^{'}$& $\mu_{ij}^{k}=-\frac{1}{V}\sum_{p\in V}v_{i}^{'}I_{ij}\dot{\theta}_{j}$\\
 \hline
\end{tabular}
\begin{tablenotes}
    \item The position vector is given by $x_{j}^{p}=x_{j}^{V_{c}}-x_{j}^{G_{p}}$, where $x_{j}^{V_{c}}$ and $x_{j}^{G_{p}}$ are the center of the RVE, and particle, respectively.
\end{tablenotes}
\end{center}
\end{table*}
The contribution of the local unbalanced force, body force, external force, contact force, boundary radius term, and kinetic fluctuation to the averaged stress, $\langle\sigma_{ij}\rangle$, can be represented by the stress tensors $\sigma_{ij}^{t}$, $\sigma_{ij}^{b}$, $\sigma_{ij}^{e}$, $\sigma_{ij}^{c}$, $\sigma_{ij}^{r}$, $\sigma_{ij}^{k}$. Contributions to averaged couple stress, $\langle \mu_{ij} \rangle$ are $\mu_{ij}^{t}$, $\mu_{ij}^{b}$, $\mu_{ij}^{e}$, $\mu_{ij}^{c}$, $\mu_{ij}^{r}$, $\mu_{ij}^{k}$. They can be calculated through the particle scale information as shown in Table.\ref{tab:table1}. 

It is worth mentioning that both the averaged stress and the averaged couple stress can be calculated from the contact pairs as previous works\cite{weber1966recherches,christoffersen1981micromechanical,bagi1996stress,babic1990stress,nicot2013definition,bardet2001asymmetry,chang2005virtual} in addition to the kinetic and boundary radius term ( which usually can be neglected in quasi-static conditions, and by choosing a large enough RVE system). We can also calculate them through the particle scale values including their positions, forces, and moments without the contact pairs. In this way, it is convenient to quantitatively analyze the effects of each force field.
\subsection{Analytical derivation of the boundary-radius-gap term} \label{section:BRG}

According to the analytical derivation above and the investigation of \cite{yan2019definition}, the boundary-radius-gap term is part of the mean stress of the granular medium. However, as the boundary-radius-gap term is dependent on the external surface force, which is unknown for an RVE, it cannot be directly obtained. A large enough number of particles inside the RVE is necessary to ignore the effect of boundary-radius-gap term \cite{guo2014coupled,nicot2013definition}. In this section, we deduce the boundary-radius-gap term from the microscopic information of each particle inside the RVE. 
In this part, we focus on the dense quasi-static granular assembly, hence the kinetic effects are ignored.

\begin{figure}
	\centering
	\includegraphics[scale=0.7]{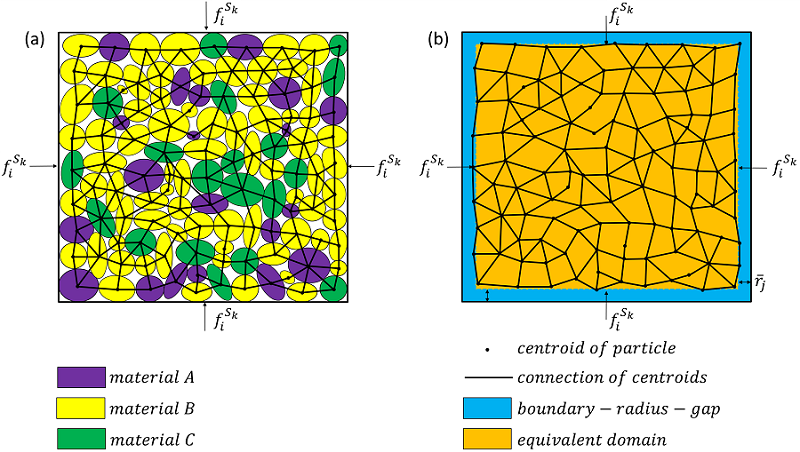}
	\caption{Transformation from particle assembly to equivalent continuum domain: (a) the particle assembly distribution, (b) the equivalent domain. }
	\label{averagevolume}
\end{figure}

First, we decompose the boundary-radius-gap term into six parts, which represent contributions from six external surfaces of a cubic RVE.
\begin{equation}
\begin{split}
\frac{1}{V}\sum_{p \in \partial V}^{}f_{i}^{e,p}r_{j}^{p}=&\frac{1}{V}\sum_{p \in S_{1}}^{}f_{i}^{e,p}r_{j}^{p}+\frac{1}{V}\sum_{p \in S_{2}}^{}f_{i}^{e,p}r_{j}^{p}+\frac{1}{V}\sum_{p \in S_{3}}^{}f_{i}^{e,p}r_{j}^{p}+\\
&\frac{1}{V}\sum_{p \in S_{4}}^{}f_{i}^{e,p}r_{j}^{p}+\frac{1}{V}\sum_{p \in S_{5}}^{}f_{i}^{e,p}r_{j}^{p}+\frac{1}{V}\sum_{p \in S_{6}}^{}f_{i}^{e,p}r_{j}^{p}, \label{equa:brg}
\end{split}
\end{equation}
where $S_{k}$ is the $\textit{k}$th external surface. In the granular assembly, the boundary-radius-gap vectors of each particle are assumed to be orthogonal to the external surface, and their magnitude is equal to the average radius $\parallel {r}_{j}^p\parallel=\bar{r}=\frac{1}{N}\sum_{p \in S_{k}}^{N}R$ of the boundary particles, where $R$ is the radius of each particle, then Eq.\ref{equa:brg} can be given by
\begin{equation}
\begin{split}
\frac{1}{V}\sum_{p \in \partial V}^{}f_{i}^{e,p}r_{j}^{p}\approx \frac{1}{V}\sum_{k=1}^{6}{r}_{j}^{S_{k}}\sum_{p \in S_{k}}^{}f_{i}^{e,p}
\label{equa:brg1}
\end{split}
\end{equation}
where the magnitude of $r_{j}^{S_{k}}$ is $\bar{r}$, its direction is normally outward to the $\textit{k}$th surface and noting that $\sum_{p \in S_{k}}^{}f_{i}^{e,p}=f_{i}^{S_{k}}$, the boundary-radius-gap term is
\begin{equation}
\frac{1}{V}\sum_{p \in \partial V}^{}f_{i}^{e,p}r_{j}^{p}\approx \frac{1}{V}\sum_{k=1}^{6}f_{i}^{S_{k}}{r}_{j}^{S_{k}}. \label{equa:brg2}
\end{equation}
where $f_{i}^{S_{k}}$ is the total force on the $\textit{i}$th surface. Taking advantage of Cauchy stress, the following relation follows
\begin{equation}
f_{i}^{S_{k}}=\sigma_{ij}n_{j}^{k} \cdot S_{k}, \label{equa:extf}
\end{equation}
where $n_{j}^{k}$ is the normal outward vector of $\textit{i}$th surface. To obtain $f_{i}^{S_{k}}$, we transform the discrete system into a continuous system with average density $\bar{\rho}={\sum_{p \in V}^{}\rho_{p}V_{p}}/{V_{e}}$, average unbalanced force density $\bar{\lambda}_{i}={\sum_{p \in V}^{}\rho_{p}V_{p}\ddot{x}_{i}^{p}}/{V_{e}}$ and the same external force $f_{i}^{S_{k}}$ and body force density $\bar{\gamma}_{i}={\sum_{p \in V_{e}}^{}f_{i}^{b,p}}/{V_{e}}$ as shown in Fig. \ref{averagevolume}, where $V_{p}$, $\rho_{p}$, $\ddot{x}_{i}^{p}$ are the volume, density, and acceleration of each particle respectively, and $V_{e}=(L_{x}-2\bar{r})\times(L_{y}-2\bar{r})\times(L_{z}-2\bar{r})$ is the volume of the equivalent domain, where $L_{x}$, $L_{y}$, and $L_{z}$ are the side lengths of the RVE in each direction.

The momentum equation of the equivalent domain holds
\begin{equation}
\frac{\partial \sigma_{ij}}{\partial x_{j}}+\bar{\gamma}_{i}=\bar{\lambda}_{i}. \label{equa:momentum}
\end{equation}
The Cauchy equation is established based on the condition that the stress tensor $\sigma_{ij}$ is continuous and differentiable at the position $x_{j}$, then the stress tensor of the equivalent domain is assumed to be a linear function of $x_{j}$, which can be expressed as
\begin{equation}
\sigma_{ij}=(\bar{\lambda}_{i}-\bar{\gamma}_{i})x_{j}+\beta_{ij}. \label{equa:func}
\end{equation}
When the position $x_{j}$ is at the geometric center ($x_{j}=0$), the stress is the average stress of the equivalent domain, hence $\beta_{ij} = \langle \sigma_{ij} \rangle_{V} $. Then the stress tensor is
\begin{equation}
\sigma_{ij}=(\bar{\lambda}_{i}-\bar{\gamma}_{i})x_{j}+\langle \sigma_{ij}\rangle_{V}. \label{equa:funct}
\end{equation}
The averaged stress tensor $\langle \sigma \rangle_{V}$ for the equivalent domain can be derived using the following expression
\begin{equation}
\langle \sigma_{ij} \rangle_{V}=\frac{1}{V_{e}} \int_{V_{e}}^{}\sigma_{ij}dV_{e}, \label{equa:oae}
\end{equation}
where $V_{e}$ is the equivalent domain. Noting that $\sigma_{ij}=\sigma_{ik}\delta_{kj}=\sigma_{ik}\frac{\partial x_{j}}{\partial x_{k}}$, where $\delta_{kj}$ is the Kronecker Delta, then Eq.\ref{equa:oae} can be rewritten as

\begin{equation}
\langle \sigma_{ij} \rangle_{V}=\frac{1}{V_{e}} \int_{V_{e}}^{}\sigma_{ik} \frac{\partial x_{j}}{\partial x_{k}}dV_{e}. \label{equa:kro}
\end{equation}
According to the Gauss theorem,
\begin{equation}
\int_{S^{e}}^{}\sigma_{ik} x_{j} n_{k}dS^{e}=\int_{V_{e}}^{}\frac{\partial \sigma_{ik}}{\partial x_{k}} x_{j}dV_{e}+\int_{V_{e}}^{}\sigma_{ik}\frac{\partial x_{j}}{\partial x_{k}} dV_{e}, \label{equa:Gauss}
\end{equation}
where $n_{k}$ is the normal outward vector of the boundary surface $S^{e}$ of the equivalent domain. Then, Eq.\ref{equa:kro} can be rewritten as
\begin{equation}
\langle \sigma_{ij} \rangle_{V}=\frac{1}{V_{e}}\int_{S^{e}}^{}\sigma_{ik}x_{j}n_{k}dS-\frac{1}{V_{e}}\int_{V_{e}}^{}\frac{\partial \sigma_{ik}}{\partial x_{k}} x_{j}dV_{e}. \label{equa:stress}
\end{equation}

Since the external force at each point of the boundary ($\partial V$) is $f_{i}^{e}=\sigma_{ij}n_{j}$, by combining with Eq.\ref{equa:momentum}, the expression of Eq.\ref{equa:stress} is then rewritten as
\begin{equation}
\langle \sigma_{ij} \rangle_{V}=\frac{1}{V_{e}}\int_{S^{e}}^{}f_{i}^{e}x_{j}dS-\frac{1}{V_{e}}\int_{V_{e}}^{}(\bar{\lambda}_{i} - \bar{\gamma_{i}})x_{j}dV_{e}. \label{equ:contimod}
\end{equation}
In the equivalent domain, the total external surface force is the same with the granular assembly, then the first term of the right-hand side of Eq.\ref{equ:contimod} is 
\begin{equation}
\frac{1}{V_{e}}\int_{S^{e}}^{}f_{i}^{e}x_{j}dS^{e}=\frac{1}{V_{e}}\sum_{p \in \partial V_{e}} f_{i}^{e,p} x_{j}^{p}.
\end{equation}
As illustrated in Section.\ref{Section:DES}, the external force subjected by particles inside the volume is zero, so we can obtain
\begin{equation}
\frac{1}{V_{e}}\sum_{p \in \partial V_{e}} f_{i}^{e,p} x_{j}^{p}=\frac{1}{V_{e}}\sum_{p \in \partial V_{e}} f_{i}^{e,p} x_{j}^{p}+\frac{1}{V_{e}}\sum_{p \in V_{e}-\partial V_{e}} f_{i}^{e,p} x_{j}^{p}=\frac{1}{V_{e}}\sum_{p \in  V_{e}} f_{i}^{e,p} x_{j}^{p}.
\end{equation}
According to the force analysis, as proposed in Eq.\ref{froceana}, the first term of the right-hand side of Eq.\ref{equ:contimod} is
\begin{equation}
\frac{1}{V_{e}}\int_{S}^{}f_{i}^{e}x_{j}dS=\frac{1}{V_{e}}\sum_{p \in V_{e}} f_{i}^{t,p}x_{j}^{p}-\frac{1}{V_{e}}\sum_{p \in V_{e}} \sum_{q \in V_{e}} f_{i}^{q,p}x_{j}^{p}-\frac{1}{V_{e}}\sum_{p \in V_{e}} f_{i}^{b,p}x_{j}^{p}.
\end{equation}
Since the average unbalanced force density $\bar{\lambda}_{i}$, and the volume body force density are constants, the second term of the right-hand side of Eq.\ref{equ:contimod} is
\begin{equation}
-\frac{1}{V_{e}}\int_{V_{e}}^{}(\bar{\lambda}_{i} - \bar{\gamma_{i}})x_{j}dV_{e}=-\frac{1}{V_{e}}(\bar{\lambda}_{i} - \bar{\gamma_{i}})\int_{V_{e}}^{}x_{j}dV_{e}. \label{equa:second}
\end{equation}
Noting that $x_{j}=x_{j}^{V_{c}}-x_{j}^{G_{p}}$ is the vector from particle mass center position $x_{j}^{G_{p}}$ to the center of the equivalent domain, $x_{j}^{V_{c}}$.  $\frac{1}{V_{e}}\int_{V_{e}}^{}x_{j}dV_{e}=\frac{1}{V_{e}}\int_{V_{e}}^{}(x_{j}^{V_{c}}-x_{j}^{G_{p}})dV_{e}=0$, hence the second term of Eq.\ref{equ:contimod} is zero,
the averaged stress tensor of the equivalent domain is
\begin{equation}
\langle \sigma_{ij}\rangle_{V} = \frac{1}{V_{e}}\sum_{p \in V_{e}} f_{i}^{t,p}x_{j}^{p}-\frac{1}{V_{e}}\sum_{p \in V_{e}} \sum_{q \in V_{e}} f_{i}^{q,p}x_{j}^{p}-\frac{1}{V_{e}}\sum_{p \in V_{e}}f_{i}^{b,p}x_{j}^{p}.
\end{equation}
Substituting the contact term with Eq.\ref{equa:lc}, the averaged stress tensor of the equivalent domain can be derived as
\begin{equation}
\langle \sigma_{ij}\rangle_{V}=\frac{1}{V_{e}}\sum_{p \in V_{e}}f_{i}^{t,p}x_{j}^{p}+\frac{1}{V_{e}} \sum_{c=1}^{N_{c}} f_{i}^{c}l_{j}^{c}-\frac{1}{V_{e}}\sum_{p \in V_{e}}f_{i}^{b,p}x_{j}^{p}.
\label{equa:geometric}
\end{equation}
It is worth noting that the averaged stress tensor of the equivalent domain is different from the work of Weber, here $\langle \sigma_{ij}\rangle_{V}$ is not the averaged stress tensor of a discrete system. As the deduction above, the stress field of the equivalent domain is obtained and combined with Eq.\ref{equa:brg2}, Eq.\ref{equa:extf}, and Eq.\ref{equa:funct}, the boundary-radius-gap term can be obtained in each RVE.

\section{Numerical inspection from numerical simulations} \label{sectionnum}
The purpose of this section is to verify the previous derivation through two tests, which are the triaxial compression tests in dry and submerged conditions using numerical simulations. These simulations are performed using DEM and LBM. The details of the methods are introduced in Section \ref{section:DEM} and Section \ref{section:LBM}. In Section \ref{section:dry}, the DEM method is used to investigate the triaxial test of dry granular materials and the contribution of each component of the stress tensor with the presence of gravity. In section \ref{section:SUB}, LBM coupled with the DEM method is used to investigate the triaxial test of submerged granular materials and the contribution of each component of the stress tensor with the presence of both gravity and hydrodynamic forces. 
In the triaxial tests, the simulations are implemented in the quasi-static condition to remove the kinetic effects.
\subsection{Contact model} \label{section:DEM}
\subsubsection{Normal inter-particle collisions}
DEM is used to simulate the particle movement in this study because the simulation data contains many quantities that are difficult to obtain from experiments, such as the transient contact force and the individual particle trajectory. These quantities are significant to investigate granular flows. Here sphere particles were used to implement our tests. When two particles contact with each other in a DEM simulation, the overlap ${\delta_{n}}$ of them is calculated as
\begin{equation}
\delta_{n}=r_{a}+r_{b}-r_{ab},
\end{equation}
where $r_{a}$ and $r_{b}$ are the radii of two contact particles, $r_{ab}$ is the distance between the contact particle centers.
We implement a Hookean contact law with energy dissipation so that the normal contact force $\boldsymbol{F_{n}}$ is given by \cite{cundall1979discrete}
\begin{equation}
\boldsymbol{F_{n}}=K_{n}\delta_{n}\boldsymbol{n}-c_{n}{\Delta}\boldsymbol{u_{n}},
\end{equation}
\begin{equation}
K_{n}=\frac{k_{n}^{a}k_{n}^{b}}{k_{n}^{a}+k_{n}^{b}},
\end{equation}
where $K_{n}$ is the effective normal stiffness, which is calculated by the normal stiffness of the contact particles $k_{n}^{a}$, $k_{n}^{b}$, $\boldsymbol{n}$ is the normal unit vector pointing from the center of particle $a$ to center of particle $b$, and ${{\Delta}\boldsymbol{u_{n}}}$ is the relative normal velocity. Here the normal viscous coefficient $c_{n}$ can be obtained using the following equations
\begin{equation}
c_{n}=\varrho_{n}\sqrt{2\bar{m}K_{n}},
\end{equation}
\begin{equation}
\varrho_{n}=\frac{-\ln e_{n}}{\sqrt{\pi^2+(\ln e_{n})^2}},
\end{equation}
where $\varrho_{n}$ is the damping ratio, $e_{n}$ is the restitution coefficient, and $\bar{m}$ is the equivalent mass which is given by
\begin{equation}
\bar{m}=\frac{m_{a}m_{b}}{m_{a}+m_{b}},
\end{equation}
where $m_{a}$ and $m_{b}$ are the masses of two contacting particles.
\subsubsection{Tangential inter-particle collisions}
The tangential contact force $\boldsymbol{F_{t}}$ is given by \cite{babic1990stress}

\begin{equation}
\boldsymbol{F_{t}}=\textrm{min}(\vert{K_{t}\boldsymbol{x_{t}}}\vert,\vert{\mu}\boldsymbol{F_{n}}\vert)\frac{\boldsymbol{x_{t}}}{\vert{x_{t}}\vert}, \label{con:Ft}
\end{equation}

\begin{equation}
K_{t}=\frac{k^{a}_{t}k^{b}_{t}}{k^{a}_{t}+k^{b}_{t}},
\end{equation}
where $\boldsymbol{x_{t}}$ is the displacement of the contact point in the tangential direction, $K_{t}$ is the stiffness in the tangential direction, which is calculated by the tangential stiffness of the contact particles $k_{t}^{a}$, $k_{t}^{b}$.
The torque driving the sphere rotation is given by
\begin{equation}
\boldsymbol{M_{c}} = \boldsymbol{F_{t}}\times \boldsymbol{{r}}, \label{con:T}
\end{equation}
where $\boldsymbol{{r}}$ is the vector from the contact point to the mass center. Further details can be found in \cite{galindo2015micro}.

\subsection{Lattice Boltzmann method} \label{section:LBM}
\begin{figure}
	\centering
	\includegraphics[scale=0.6]{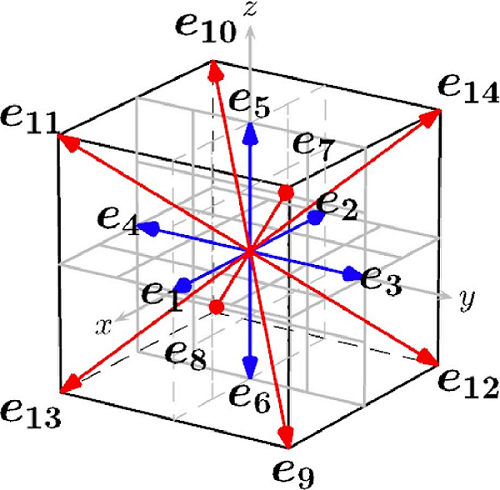}
	\caption{The D3Q15 cell showing the direction of each one of the 15 discrete velocities.\cite{galindo2013coupled}}
	\label{Lattice}
\end{figure}

LBM is widely used for simulating fluids since it can provide accurate fluid information on mesoscopic scales and has high performance on parallel architectures. It is suitable for mass-conserving problems in complex geometries such as porous media flow. Since pore-scale fluid-solid interactions dominate the submerged granular column collapse, a D3Q15 LBM is used in this work. This model is used for 3-D LBM simulation, and each cell has 15 discrete velocities, as shown in Fig. \ref{Lattice}. The 15 velocity vectors are defined as follows

\begin{equation}
\boldsymbol{e_{i}}=\left \{
\begin{array}{lr}
0,&i=0, \\
(\pm C,0,0),(0,\pm C,0),(0,0,\pm C),&i=1 \sim 6,\\
(\pm C,\pm C,\pm C),&i=7 \sim 14,
\end{array}
\right.
\end{equation}
where $C = {\delta}x/{\delta}t$ is the characteristic lattice velocity. The Chapman-Enskog expansion of the Boltzmann equation is given by
\begin{equation}
f_{i}(\boldsymbol{x}+\boldsymbol{e_{i}}\delta{t},t+\delta{t})=f_{i}(\boldsymbol{x},t)+\Omega_{col},
\end{equation}
where $\boldsymbol{x}$ is the position of the local cell, $\delta {t}$ is the time step, and $\Omega_{col}$ is the collision operator. The Bhatnagar-Gross-Krook(BGK) collision operator introduced in \cite{qian1992lattice} is used in this study, which is
\begin{equation}
\Omega_{col}=\frac{\delta{t}}{\tau}(f_{i}^{eq}-f_{i}),
\end{equation}
where $\tau$ is the characteristic relaxation time, and $f^{eq}_{i}$ is the equilibrium function given by
\begin{equation}
f_{i}^{eq}=\omega_{i}\rho_{f}\left(1+3\frac{\boldsymbol{e_{i}}\cdot\boldsymbol{u}}{C^2}+\frac{9(\boldsymbol{e_{i}}\cdot\boldsymbol{u})^2}{2C^4}-\frac{3\boldsymbol{u}^2}{2C^2}\right),
\end{equation}
where ${\delta}x$ is the lattice size, $\rho_{f}$ is the fluid density, $\boldsymbol{u}$ is the fluid flow velocity, and the weights are
\begin{equation}
\omega_{i}=\left\{
\begin{array}{lr}
2/9 &i=0, \\
1/9 &i=1 \sim 6,\\
1/72 &i=7 \sim 14.
\end{array}
\right.
\end{equation}
The kinetic viscosity $\nu$ is related to the relaxation time by
\begin{equation}
\nu=(\tau-0.5)\frac{{\delta}^2_{x}}{3\delta_{t}}.
\end{equation}
The fluid density $\rho_{f}$, and fluid flow velocity $\boldsymbol{u}$ could be determined by the zeroth and first-order moments of the distribution function:
\begin{subequations}
	\begin{align}
	&\rho_{f}(\boldsymbol{x})=\sum_{i=0}^{14}f_{i}(\boldsymbol{x}),\\
	&\boldsymbol{u}(\boldsymbol{x})=\frac{1}{\rho_{f}(\boldsymbol{x})}\sum_{i=0}^{14}f_{i}(\boldsymbol{x})\boldsymbol{e_{i}},
	\end{align}
\end{subequations}
And the pressure $p_{f}$ of the fluid is given by
\begin{equation}
p_{f}(\boldsymbol{x})=\frac{1}{3}C^{2}\rho_{f}(\boldsymbol{x}). \label{equa:p}
\end{equation}
The standard LBM is only suitable for flow at a low Reynolds number because the value of relaxation time $\tau$ cannot be too close to 0.5. In this study, the Smagorinsky subgrid turbulence model was used to simulate the fluid flow at high Reynolds numbers. The scale larger than lattice size ${\delta}x$ is defined as the filtered scale, and the smaller one is defined as the unresolved scale. LBM can be directly solved for the filtered scales. An additional relaxation time $\tau_{a}$, which is related to the turbulence viscosity $\nu_{a}$, is used to describe the influence of fluid flow at unresolved scales \cite{van2005lattice},
\begin{equation}
\tau_{total}=\tau+\tau_{a},
\end{equation}
\begin{equation}
\tau_{a}=\frac{3\delta_{t}}{{\delta}^2_{x}}\nu_{a},
\end{equation}
the turbulence viscosity $\nu_{a}$ is given by
\begin{equation}
\nu_{a}=(S_{c}{\delta}x)^2\widetilde{S},
\end{equation}
where $S_{c}$ is the Smagorinsky constant which a value is between 0.1 and 0.2, $\widetilde{S}$ is the magnitude of the filtered strain-rate tensor given by
\begin{equation}
\widetilde{S}=\frac{\sqrt{2\widetilde{Q}_{ij}\widetilde{Q}_{ij}}}{2\rho_{f} S_{c}\tau_{total}},
\end{equation}
where $\widetilde{Q}_{ij}$ is the second moment of the distribution function, which is:
\begin{equation}
\widetilde{Q}_{ij}=\sum_{k=0}^{14}e_{ki}e_{kj}(f_{k}-f^{eq}_{k}),
\end{equation}
The modified LBM introduced by \cite{feng2007coupled,feng2010combined,owen2011efficient}, which is based on the immersed boundary method\cite{noble1998lattice}, was used to simulate the fluid-solid interaction. The Lattice Boltzmann equation is modified as
\begin{equation}
f_{i}(\boldsymbol{x}+\boldsymbol{e_{i}}\delta{t})=f_{i}(\boldsymbol{x},t)+B_{n}\Omega^{S}_{i}+(1-B_{n})\left[\frac{\delta{t}}{\tau}(f^{eq}_{i}-f_{i})\right],
\end{equation}
where $B_{n}$ is a weighting function. $\Omega^{S}_{i}$ is the collision operator proposed by \cite{noble1998lattice}, which accounts for the momentum exchange between fluid and solid. The bounce-back rule is applied to the interface of the fluid and solid, and hence, $\Omega^{S}_{i}$ is given by
\begin{equation}
\Omega^{S}_{i}=f_{i}(\boldsymbol{x},t)-f^{eq}_{i}(\rho_{f},\boldsymbol{u_{s}})+f^{eq}_{i}(\rho_{f},\boldsymbol{u_{s}})-f_{i}(\boldsymbol{x},t),
\end{equation}
where $\boldsymbol{u_{s}}$ is the macroscopic velocity of the particle at cell position $\boldsymbol{x}$, which is
\begin{equation}
\boldsymbol{u_{s}}=\boldsymbol{\omega}\times(\boldsymbol{x}-\boldsymbol{x_c})+\boldsymbol{v_{c}},
\end{equation}
which depends on the solid particle velocity $\boldsymbol{v_{c}}$, angular velocity $\boldsymbol{\omega}$, and position $\boldsymbol{x_{c}}$ of the sphere's center of mass.
In this study, the weight function $B_{n}$ is:
\begin{equation}
B_{n}(\varepsilon,\tau)=\frac{\varepsilon_{n}(\tau-1/2)}{(1-\varepsilon_{n})+(\tau-1/2)},
\end{equation}
where $\varepsilon_{n}$ is the volume occupation fraction, which is given by \cite{galindo2013coupled}
\begin{equation}
\varepsilon_{n}=\frac{\sum_{e=1}^{12}l_{e}}{12\delta_{x}}
\end{equation}
where $l_{e}$ is the length of the $\textit{e}$th edge occupied by solid particles.
The total hydrodynamic force and torque over a particle covered by $n$ cells are
\begin{equation}
\boldsymbol{F}_{f}=\frac{\delta_{x}^{3}}{\delta_{t}}\sum_{n}^{}B_{n}(\sum_{i}^{}\Omega^{S}_{i}{e}_{i}),
\end{equation}
\begin{equation}
\boldsymbol{T_{f}}=\frac{\delta_{x}^{3}}{\delta_{t}}\sum_{n}^{}\left[(\boldsymbol{x_{n}}-\boldsymbol{x_{c}}){\times}B_{n}(\sum_{i}^{}\Omega^{S}_{i}{e}_{i})\right].
\end{equation}
where $\boldsymbol{x_{n}}$ is the coordinates of the ${n}$th lattice cell.

\subsection{Triaxial tests of dry granular systems in quasi-static condition}\label{section:dry}

\begin{figure}
	\centering
	\includegraphics[scale=0.6]{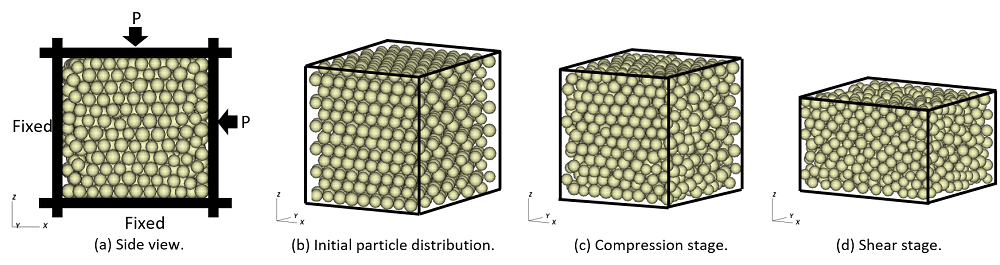}
	\caption{The triaxial test in the dry condition.}
	\label{drytest}
\end{figure}

Dry triaxial compression tests are implemented in this part in order to verify the obtained average stress formulas in the absence of hydraulic forces. The particles are initially distributed in the space without interaction, as shown in Fig. \ref{drytest}(b) in a hexagonal packing. At the initial state, the cubic container has 5 cm width, 5 cm length, and 5cm height, composed of 1188 particles. The particle radii are in the range of 0.2-0.25 cm, the frictional coefficient between particles is 0.34, the normal stiffness is $1\times10^{7} $g/s$^{2}$, the tangential stiffness is $2.5\times10^{6} $g/s$^{2}$, the restitution coefficient is 0.2, and the density of the particles $\rho_{p}$ is 5 g/cm$^{3}$. The friction between the particles and plane is assumed to be zero. During the test, the particles are subjected to the acceleration $a_{g}$(representing a constant body force akin to gravity) in the negative z-direction. To obtain more general results, we select $a_{g}=800$ cm/s$^{2}$ to represent the influence of gravity.
The test consists of two stages: First is the compression stage, where the bottom(in the lower part in the z-direction), left(in the lower part in the x-direction), and front(in the lower part in the y-direction) planes are fixed. The top(in the upper part in the x-direction), right(in the upper part in the x-direction), and back(in the upper part in the x-direction) planes are subjected to the same force($\parallel f_{i}^{l}\parallel$ = 2.25$\times$10$^{5}$ dyn) to compress the granular assembly until the volume does not change. The second is the shear stage. The plane in the x and y direction holds the same external force as the first stage, while the top plane moves downward with a constant speed (0.2 cm/s) for 5 s to ensure granular assembly is in a quasi-static condition ( which the inertial number $I=\dot{\gamma}d\sqrt{\rho_{p}/P}<1\times10^{-3}$ defined in \cite{da2005rheophysics}, where $\dot{\gamma}$ is the shear rate, $d$ is the diameter of the particle, and $P$ are the confining pressure in this study). 

\subsubsection{Validation of the total external force}
\begin{figure}
	\centering
	\includegraphics[scale=0.64]{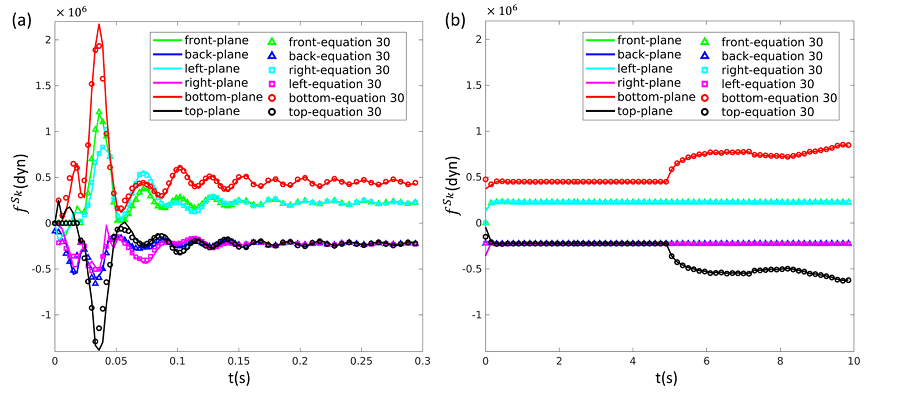}
	\caption{External force calculated using grain scale information inside the RVE and load on each plane in the triaxial test for the dry case: (a) t$<$0.3s, (b) 0$<$t$<$10s.}
	\label{externaltest}
\end{figure}

The grain scale information, such as the contact force, the gravity, the local unbalanced force, and the load on each plane, was recorded. According to the method proposed in Section \ref{section:BRG}, the body force in this test can be expressed as
\begin{equation}
f_{i}^{b,p}=w_{i}^{p},
\end{equation}
where $w_{i}^{p}$ is the gravity subjected to each particle.
Then the averaged stress tensor of the equivalent domain in Eq.\ref{equa:geometric} is transformed into
\begin{equation}
\langle \sigma_{ij}\rangle_{V}=\frac{1}{V_{e}}\sum_{p \in V_{e}}f_{i}^{t,p}x_{j}^{p}+\frac{1}{V_{e}} \sum_{c=1}^{N_{c}} f_{i}^{c}l_{j}^{c}-\frac{1}{V_{e}}\sum_{p \in V_{e}}w_{i}^{p}x_{j}^{p}.
\label{equa:dry}
\end{equation}
To validate the external force obtained using the grain scale information, the total external surface force $f_{i}^{S_{k}}$ exerted on the assembly by each plane is used to make the comparison,
\begin{equation}
f_{i}^{S_{k}}=f_{i}^{t,S_{k}}-f_{i}^{l},
\end{equation}
where $f_{i}^{t,S_{k}}$, and $f_{i}^{l}$ are the resultant force and the loading on each plane, respectively. 
As shown in Fig. \ref{externaltest}, the load on each plane and the total external force calculated using the grain scale information agree well with each other during the compression and the shear stage. At the beginning of the compression stage(0-0.3 s), particles start to contact each other. The force exerted on the boundary particles transmits to the inside particles. The total external surface force subjected by the boundary particles fluctuated with time. When the force is transmitted from the surface plane to the assembly, the local unbalanced force of the particles is large enough to generate the propagation of stress waves.
When (time 0.3-5 s) the granular assembly is in the static stage, the local unbalanced force of each particle is nearly zero. Hence the external surface forces are also stable. The surface force on the bottom plane is different from the top and lateral planes, a consequence of the body force of the granular assembly. When the top plane starts to move downward, the granular assembly starts to shear. The surface forces of the top and bottom planes increase in the first 2s; then, the surface forces subjected by boundary particles decrease; at last, the forces increase again.

\subsubsection{Validation of the mean stress}
\begin{figure}
	\centering
	\includegraphics[scale=0.33]{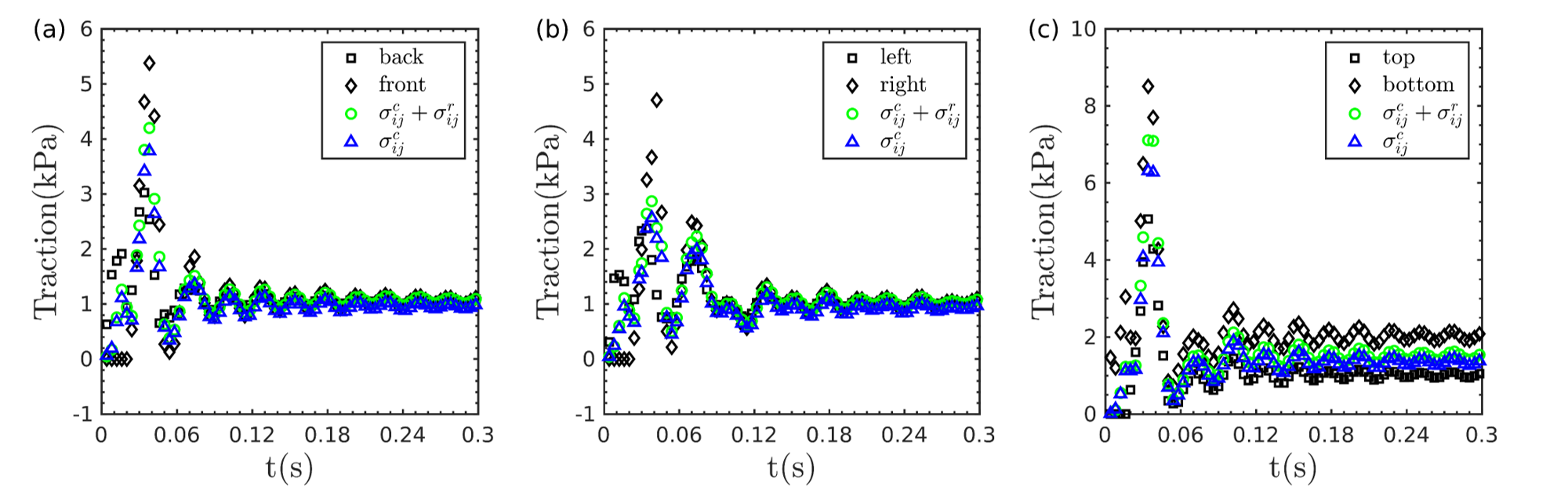}
	\caption{Traction obtained from the averaged stress tensor and pressure of each plane at the beginning of the compression stage(0$<$t$<$0.3s): (a) in the x-direction, (b) in the y-direction, (c) in the z-direction.}
	\label{stressdrys}
\end{figure}

\begin{figure}
	\centering
	\includegraphics[scale=0.33]{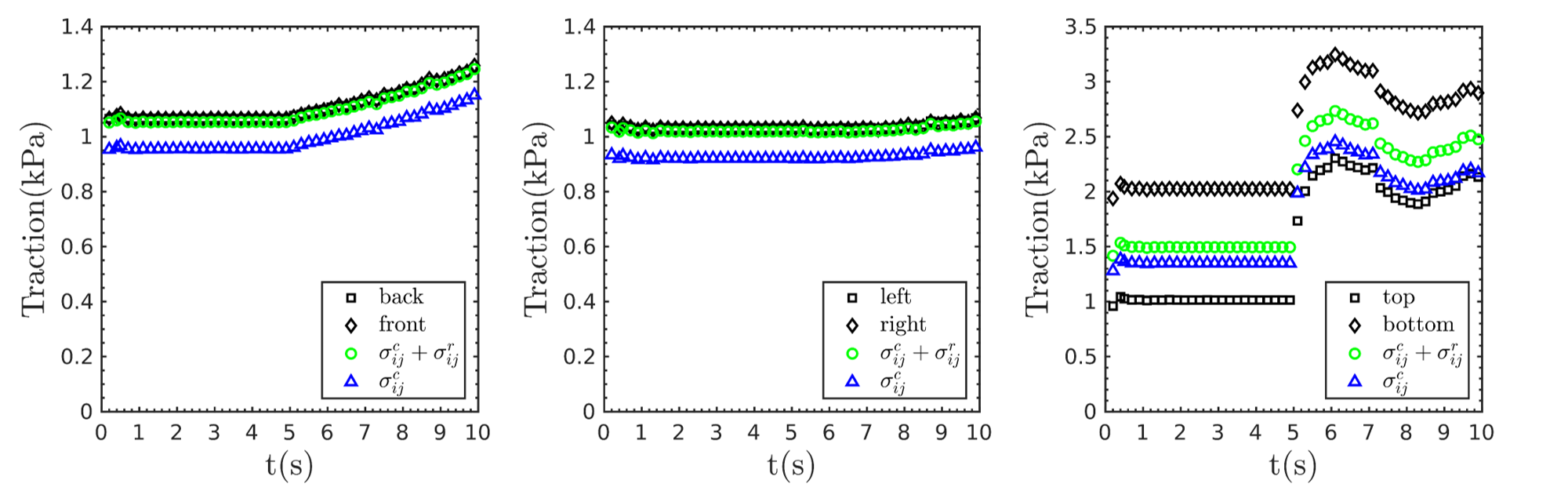}
	\caption{Traction obtained from the averaged stress tensor and pressure of each plane of the compression and shear stage (0$<$t$<$10s): (a) in the x-direction, (b) in the y-direction, (c) in the z-direction.}
	\label{stressdryl}
\end{figure}

As shown in Figs. \ref{stressdrys} and \ref{stressdryl}, traction on each plane is presented. The pressure $P_{i}^{e,S_{k}}$ on each plane is calculated using the surface force $f_{i}^{S_{k}}$ divided by area $S_{k}$, is used to validate the traction of the average stress given by
\begin{equation}
    \langle \sigma_{ij} \rangle=\sigma_{ij}^{c}+\sigma_{ij}^{r}.
    \label{equa:conponents}
\end{equation}
The kinetic fluctuation term $\sigma_{ij}^{k}$ is neglected in this quasi-static condition. 
\begin{equation}
P^{e,S_{k}}=\frac{f_{i}^{S_{k}}n_{j}^{k}}{S_{k}}, \quad  k=1 \sim 6 . \label{equa:wpre}
\end{equation} 
The traction of the average stress obtained using Eq.\ref{equa:conponents} shows good agreement with the loading on the lateral plane (along x- and y-direction). 
Due to the presence of gravity, the bottom plane is subjected to a more significant surface force, as shown in figure \ref{externaltest}, and the top plane pressure is smaller than the bottom plane pressure. Hence the average stress in the z-direction is in the middle between top pressure and bottom pressure. However, the traction of average stresses without the boundary-radius-gap term from Eq.\ref{equa:previous} is different from the plane pressure in each direction. 
The traction exhibits fluctuation, also called stress wave, at the beginning of the compression stage because of the force transmission. In the static stage, traction in the three directions is stable. The traction is similar in the x and y direction because of the same external surface force. In the z-direction, the traction is more significant than in the lateral direction.
In the shearing stage, traction in the lateral plane barely changes. In the z-direction, the average stress increase first, and then decrease, at last increase again, which is influenced by the external force subjected by the boundary particles. The match between Eq.\ref{equa:conponents} and the observed pressures in the lids show the need to include boundary-radius-gap term correction to Weber's formula.

\subsubsection{Effect of boundary-radius-gap on the mean stress}
\begin{figure}
	\centering
	\includegraphics[scale=0.6]{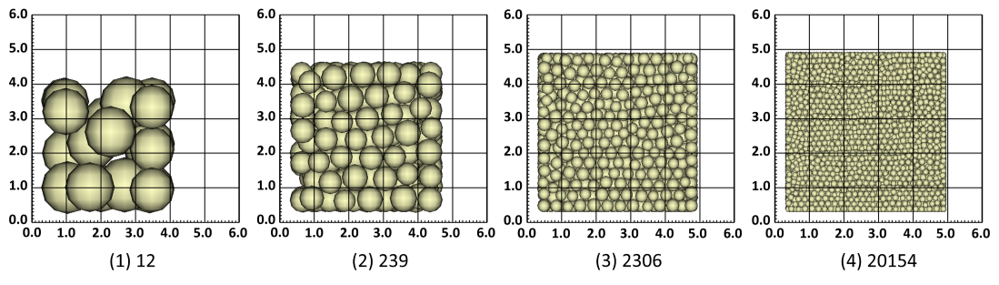}
	\caption{Different number of the particle distribution.}
	\label{numBRG}
\end{figure}

DEM simulations with different numbers of particles were performed to investigate the influence of the boundary-radius-gap term and its relationship with the system size. The number of particles shown in Fig. \ref{numBRG} is 12, 239, 2036, and 20154, respectively. Each particle assembly configuration was implemented in different conditions, respectively, as shown in Table.\ref{Rsstudy}. The size effect factor, $R_{s}$, was used to evaluate this influence,
\begin{subequations}
	\begin{align}
	&R_{s1}=\frac{L_{x}}{\bar{r}},\\
	&R_{s2}=\frac{L_{y}}{\bar{r}},\\
	&R_{s3}=\frac{L_{z}}{\bar{r}}.
	\end{align}
\end{subequations}
The relative error $\zeta_{i}$($i=1,2,3$) is defined as the contribution of boundary-radius-gap divided by the average stress in the principal direction. 
\begin{subequations}
	\begin{align}
	&\zeta_{1} = \frac{\sigma_{xx}^{r}}{\langle \sigma_{xx} \rangle},\\
	&\zeta_{2} = \frac{\sigma_{yy}^{r}}{\langle \sigma_{yy} \rangle},\\
	&\zeta_{3} = \frac{\sigma_{zz}^{r}}{\langle \sigma_{zz} \rangle}.
	\end{align}
\end{subequations}

\begin{table}
	\begin{center}
		\caption{Details of parameters used for each simulation}
		\begin{tabular}{cccc}
		\hline
			case  & particle-particle friction coefficient   &  $a_{g}$(cm/s$^{2}$) & $\parallel f_{i}^{l}\parallel$(N)  \\
			\hline
			1&0.25&800&2.25$\times$10$^{6}$\\
			2&0.25&800&4.5$\times$10$^{5}$\\
			3&0.25&500&2.25$\times$10$^{6}$\\
			4&0.5&800&2.25$\times$10$^{6}$\\
			\hline
		\end{tabular}
		\label{Rsstudy}
	\end{center}
\end{table}

\begin{figure}
	\centering
	\includegraphics[scale=0.4]{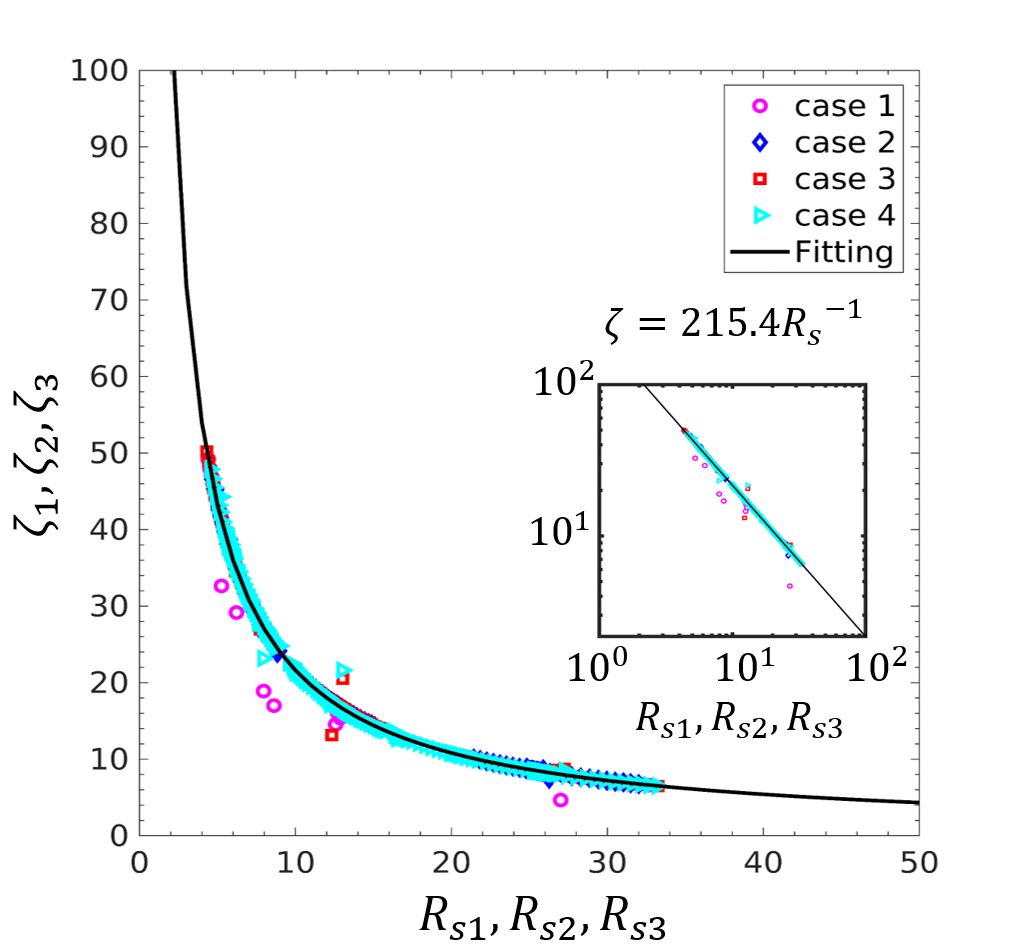}
	\caption{Relative error of the averaged stress tensor with different size effect factors in different conditions}
	\label{Rs}
\end{figure}

The results show that the relative error only depends on $R_{s}$. As we change the load of the surface force, gravity, and friction coefficient, the relation between the relative error and the size effect factor always obeys a power law curve, as shown in Fig. \ref{Rs}. The relative error decreases with the increase of $R_{s}$.
According to this study, if we use weber's formula directly, with no boundary-radius-gap correction term like in the previous works \cite{weber1966recherches,christoffersen1981micromechanical,bagi1996stress,nicot2013definition,fortin2003construction,bardet2001asymmetry}, $R_{s}$ should be larger than 30 to obtain the averaged stress tensor with an error below 10$\%$. Because of the power-law relation between $R_{s}$ and the relative error, a high $R_{s}=100$ would be necessary to obtain the high accuracy averaged stress tensor in which the relative error is less than 1$\%$. This limitation reduces the robustness of the REV method. Previous studies such as in \cite{nicot2013definition} used to select the REV with $R_{s}$ in the range of 5-30, which leads to a significant error (nearly 10$\%$-40$\%$) of the calculation of the averaged stress tensor.
At the same time, higher $R_{s}$ means that more particles are inside RVE, which increases the computational cost. The particle size might also influence the mechanism of the assembly deformation, as already presented in the granular collapse study. For instance, it is a known fact that the relative size of the particle size to the system size has a strong influence on the morphology of granular column collapses such in \cite{cabrera2019granular,ladd1993numerical,man2020finite}. 

\subsection{Triaxial tests in submerged quasi-static condition}\label{section:SUB}

\begin{figure}
	\centering
	\includegraphics[scale=0.1]{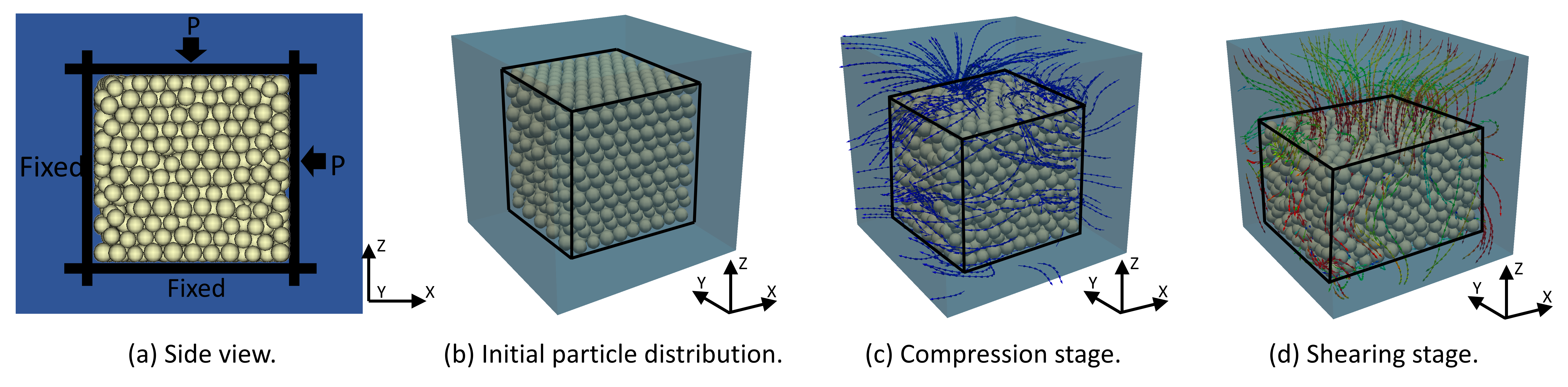}
	\caption{The triaxial test in submerged condition.}
	\label{subtest}
\end{figure}

In this section, LBM coupled with DEM was used to simulate the drained and undrained triaxial compression tests of saturated granular materials, as shown in Fig. \ref{subtest}(a). Tests in this part consist of two stages: compression and shear. 
At the beginning of the drained test, the same particle distribution as the dry case in section.\ref{section:dry} is submerged in fluid as shown in Fig. \ref{subtest}(b), and then the same force (2.25$\times$10$^{5}$ dyn) as the dry test is applied to the top, back, and right planes, fixing the bottom, front, and left the planes. All the planes are perfectly permeable so that the fluid can go through them without any resistance. With forces applied on the plane, the granular assembly was compressed till the volume does not change anymore (0.3 - 5 s), and the fluid flow velocity becomes nearly zero. At this point, the granular assembly is in a static stage. After the compression stage, the top plane moves downward with a speed of 0.2 cm/s for 5 s to ensure the granular assembly is in the same quasi-static condition as in the dry case. As shown in Fig. \ref{subtest}(d), the fluid will go through the granular assembly from the top and get out of the assembly from the bottom and lateral direction. 
The particle material parameters are the same as the dry test. The effective gravitational acceleration $a_{g}(\rho_{p}-\rho_{l})/\rho_{p}$ is applied to each particle, where $\rho_{l}$=1.0 g/cm$^{3}$ is the density of the fluid. The dynamic viscosity of the fluid is 5 g/(s$\cdot$ cm), and the grid of the LBM is 0.05 cm so that the lattice resolution $N \approx 10$ (10 LBM grids per particle diameter) to ensure sufficient accuracy \cite{ladd1993numerical}. The speed of sound ($C$) is 1000 cm/s which is much larger than the shear velocity. 

\subsubsection{Validation of the total external force}
\begin{figure}
	\centering
	\includegraphics[scale=0.64]{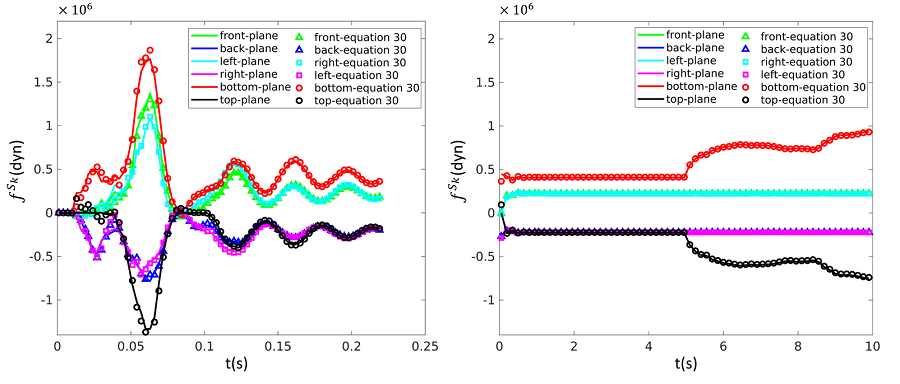}
	\caption{External force calculated using grain scale information inside the RVE and load on each plane in the triaxial test for the submerged case: (a) t$<$0.3s, (b) 0$<$t$<$10s.}
	\label{BRGFsubsl}
\end{figure}

Due to the presence of the gravity $w_{i}^{p}$ and the hydrodynamic force $f_{i}^{h,p}$, the body force should be expressed as
\begin{equation}
f_{i}^{b,p}=w_{i}^{p}+f_{i}^{h,p}.
\end{equation}
Then the averaged stress tensor of Eq.\ref{equa:geometric} should be transformed into
\begin{equation}
\langle \sigma_{ij}\rangle_{V}=\frac{1}{V_{e}}\sum_{p \in V_{e}}f_{i}^{t,p}x_{j}^{p}+\frac{1}{V_{e}} \sum_{c=1}^{N_{c}} f_{i}^{c}l_{j}^{c}-\frac{1}{V_{e}}\sum_{p \in V_{e}}(w_{i}^{p}+f_{i}^{h,p})x_{j}^{p}.
\label{equa:submerged}
\end{equation}
As shown in Fig. \ref{BRGFsubsl}, the surface force calculated using Eq.\ref{equa:extf} shows good agreement with the load. It means the proposed method is also suitable for granular assembly subjected to hydrodynamic force fields. 
\subsubsection{Validation of the mean stress}

The components of the stress tensor for the submerged case in a quasi-static condition can be expressed as
\begin{equation}
\langle\sigma_{ij}\rangle=\sigma_{ij}^{t}+\sigma_{ij}^{e}+\sigma_{ij}^{r}+\sigma_{ij}^{b}=\sigma_{ij}^{t}+\sigma_{ij}^{e}+\sigma_{ij}^{r}+\sigma_{ij}^{w}+\sigma_{ij}^{h}, \label{equa:subs}
\end{equation}
where $\sigma_{ij}^{h}$ is the contribution of the hydrodynamic force term including in the mean stress,
\begin{equation}
\sigma_{ij}^{h}=\frac{1}{V}\sum_{p \in V}^{} f_{i}^{h,p}x_{j}^{p}, \label{equa:hydro}
\end{equation}
taking advantage of the LBM, the hydrodynamic force $f_{i}^{h,p}$ subjected to each particle is obtained from the momentum exchange directly without any assumption (such as the law of drag force, lubrication force, etc.). It is worth noting that $\langle\sigma_{ij}\rangle$ obtained in the submerged case is the stress of the solid skeleton, which is also called the effective stress in \cite{terzaghi1943theoretical}. Hence, $\sigma_{ij}^{h}$ represents the real effects of fluid on the granular assembly.
\begin{figure}
	\centering
	\includegraphics[scale=0.33]{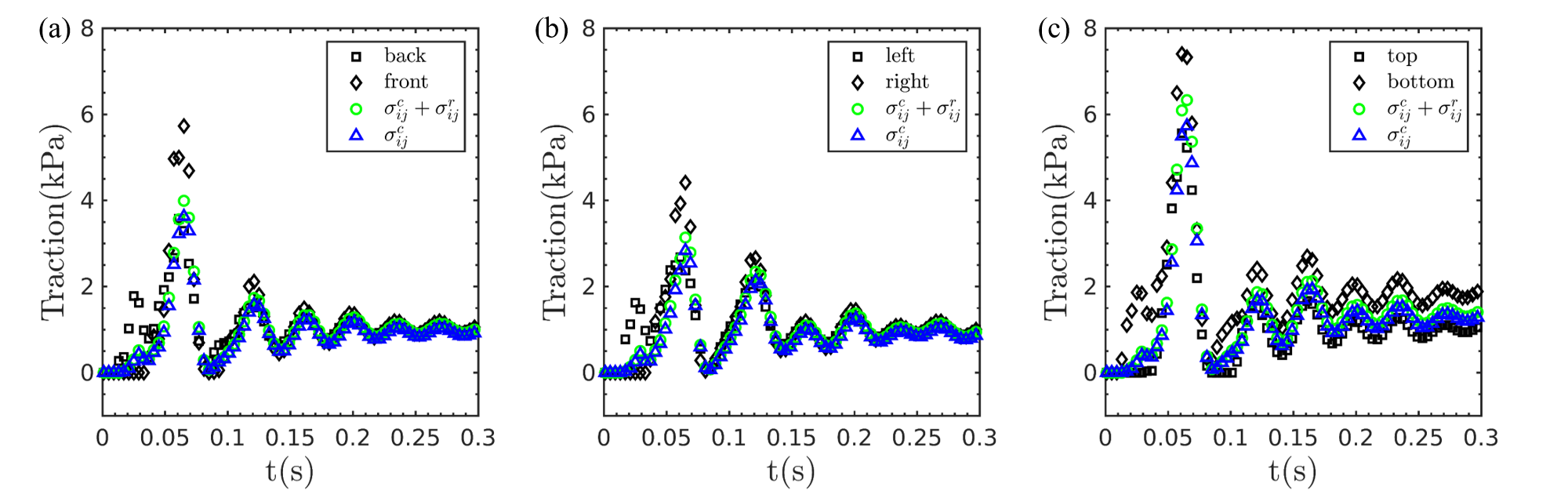}
	\caption{Traction obtained from the averaged stress tensor and pressure of each plane at the beginning of the compression stage(0$<$t$<$0.3s): (a) in the x-direction, (b) in the y-direction, (c) in the z-direction.}
	\label{stresssubs}
\end{figure}

\begin{figure}
	\centering
	\includegraphics[scale=0.33]{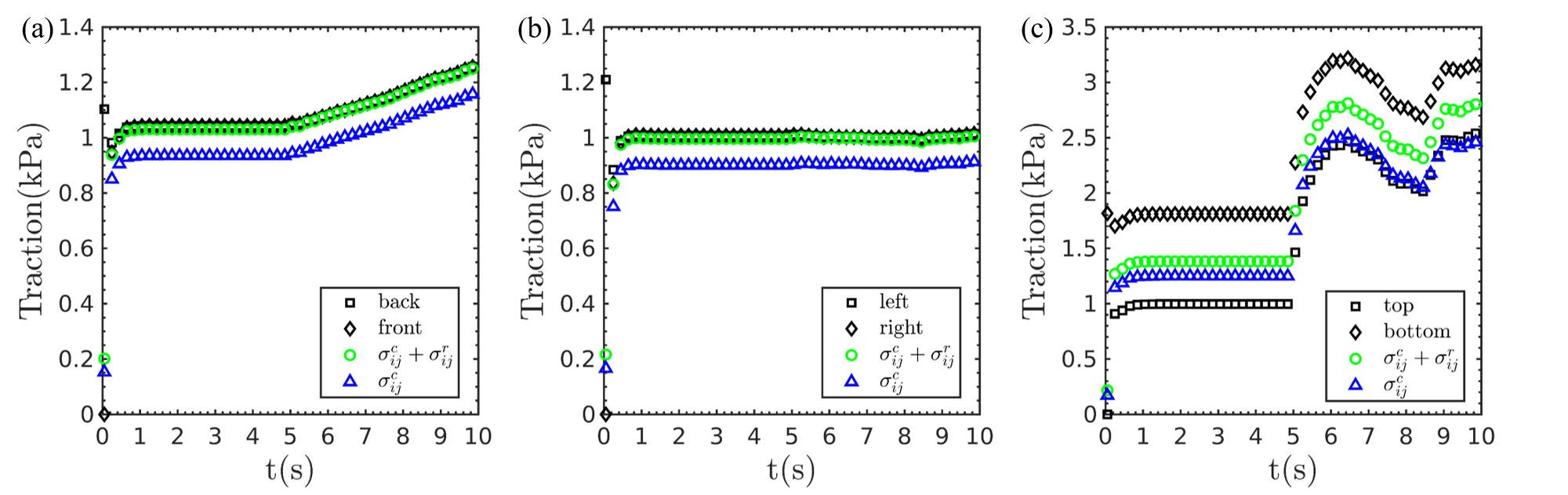}
	\caption{Traction obtained from the averaged stress tensor and pressure of each plane on the compression and shear stage (0$<$t$<$10s): (a) in the x-direction, (b) in the y-direction, (c) in the z-direction.}
	\label{stresssubl}
\end{figure}
At the beginning of the compression stage(0 - 0.3 s), as shown in Fig. \ref{stresssubs}, both the plane pressure and the traction along each direction of averaged stress tensor obtained using Eq.\ref{equa:conponents} and Eq.\ref{equa:previous} oscillate. As time goes by, the traction becomes stable, as shown in Fig. \ref{stresssubl}. 
The lateral traction of average stress obtained using Eq.\ref{equa:conponents} shows good agreement with the pressure of the wall in the compression and shear stages, while the traction obtained from the formula of Weber\cite{weber1966recherches} shows a difference with the pressure of the wall. In the z-direction, the traction obtained using the proposed method is close to the average pressure of the top and bottom walls. The difference in pressure between the bottom and top planes is due to gravity. The difference between the traction obtained using Eq.\ref{equa:conponents} and Eq.\ref{equa:previous} is larger in the shear stage compare with the compression stage. Hence in both stages, the averaged stress tensor obtained using Eq.\ref{equa:conponents} is more accurate.

\subsubsection{Contribution of each component on the average stress}
\begin{figure}
	\centering
	\includegraphics[scale=0.13]{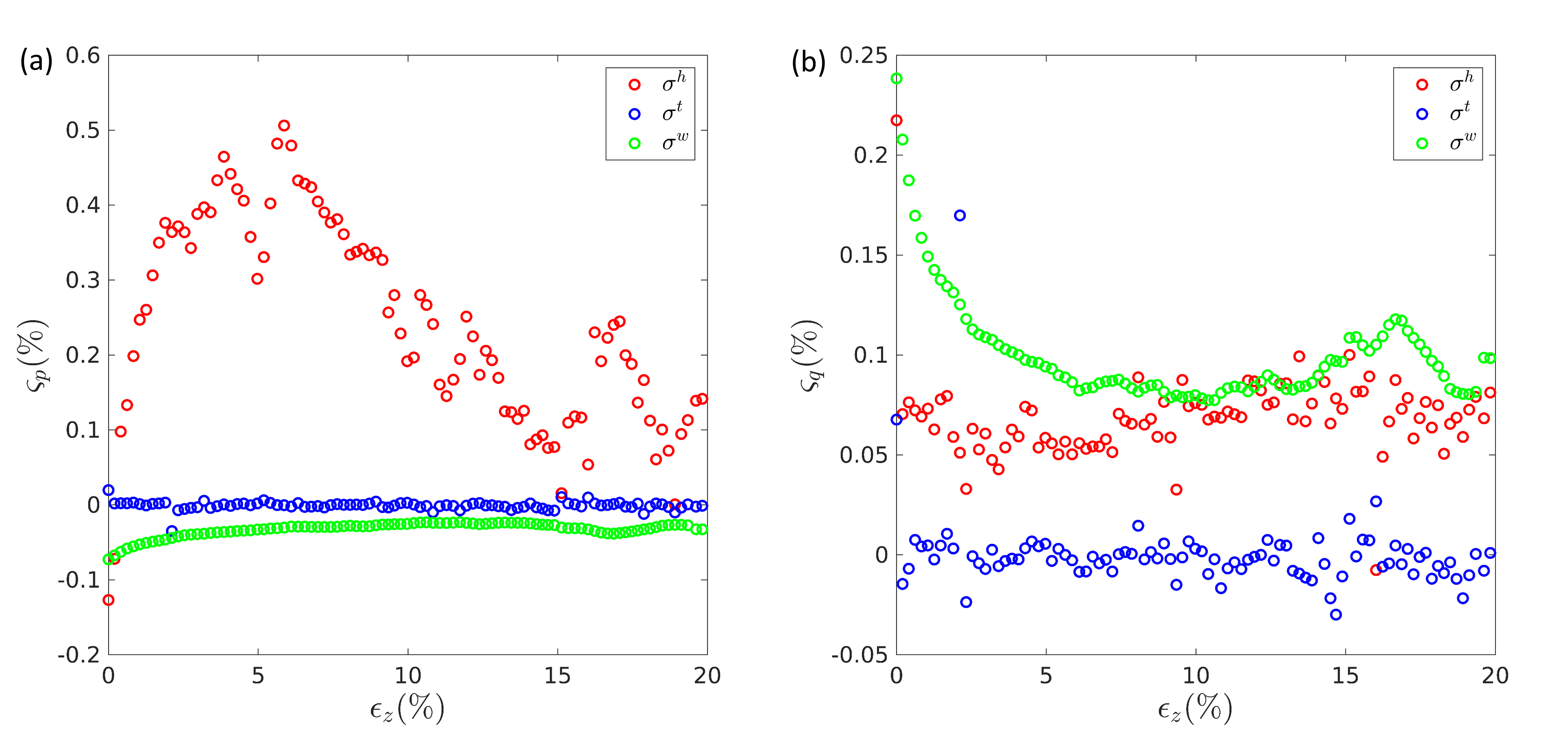}
	\caption{Variation of $\varsigma_{p}$ and $\varsigma_{q}$ (contribution of $\sigma_{ij}^{h}$, $\sigma_{ij}^{t}$, and $\sigma_{ij}^{w}$ to the mean principal stress and deviatoric stress) in terms of the axial strain $\epsilon_{z}$ during the shear stage.}
	\label{puA}
\end{figure}

Since $\sigma_{ij}^{h}$, $\sigma_{ij}^{t}$, and $\sigma_{ij}^{w}$ have the contribution to the average stress in both normal part and shear part, the following quantities are introduced to illustrate these effects,
\begin{subequations}
	\begin{align}
	&\varsigma_{p}^{h}=\frac{p-p_{\Delta h}}{p},\\
	&\varsigma_{p}^{w}=\frac{p-p_{\Delta w}}{p},\\
	&\varsigma_{p}^{t}=\frac{p-p_{\Delta t}}{p},\\
	&\varsigma_{q}^{h}=\frac{q-q_{\Delta h}}{q},\\
	&\varsigma_{q}^{w}=\frac{q-q_{\Delta w}}{q},\\
	&\varsigma_{q}^{t}=\frac{q-q_{\Delta t}}{q},
	\end{align} 
\end{subequations}
where $p$, $p_{\Delta h}$, $p_{\Delta w}$, $p_{\Delta t}$ are the mean principal stress of the following stresses: $\langle\sigma_{ij}\rangle$, $\langle \sigma_{ij} \rangle-\sigma_{ij}^{h}$, $\langle\sigma_{ij}\rangle-\sigma_{ij}^{w}$, and $\langle \sigma_{ij} \rangle-\sigma_{ij}^{t}$ respectively; $q$, $q_{\Delta h}$, $q_{\Delta w}$, $q_{\Delta t}$ is the deviatoric stress of the stresses: $\langle\sigma_{ij}\rangle$, $\langle\sigma_{ij}\rangle-\sigma_{ij}^{h}$, $\langle\sigma_{ij}\rangle-\sigma_{ij}^{w}$, and $\langle \sigma_{ij} \rangle-\sigma_{ij}^{t}$ respectively.
The mean principal stress $p$ and deviatoric stress $q$ of each stress are given by
\begin{subequations}
	\begin{align}
	& p =\frac{(\sigma_{1} + \sigma_{2} + \sigma_{1})}{3}\\
	& q =\frac{\sqrt{(\sigma_{1} - \sigma_{2} )^{2}+( \sigma_{2} -\sigma_{3} )^{2}+( \sigma_{3} - \sigma_{1} )^{2}}}{\sqrt{2}}
	\end{align}
\end{subequations}
where $\sigma_{1}$, $\sigma_{2}$, and $\sigma_{3}$ are the principal stresses.
As shown in Fig. \ref{puA}, the hydrodynamic force, unbalanced force, and gravity term show less influence on the averaged stress tensor with a value for $ \varsigma $ below 1$\%$ in terms of axial strain $\epsilon_{z}$. At the beginning of the shear stage, the influence of the hydrodynamic force term is negative. After a short time of shearing, it changes into positive, which means the fluid promotes granular dilation first and then compresses the granular assembly. The gravity shows a larger contribution to the deviatoric stress at the beginning of the shear stage and decreases time since, as seen in Fig. \ref{subtest}(d), the height of the granular column decreases. The local unbalanced force term shows nearly no influence on the structure of the granular assembly, signalling perfect quasi-static conditions. 

\begin{figure}
	\centering
	\includegraphics[scale=0.4]{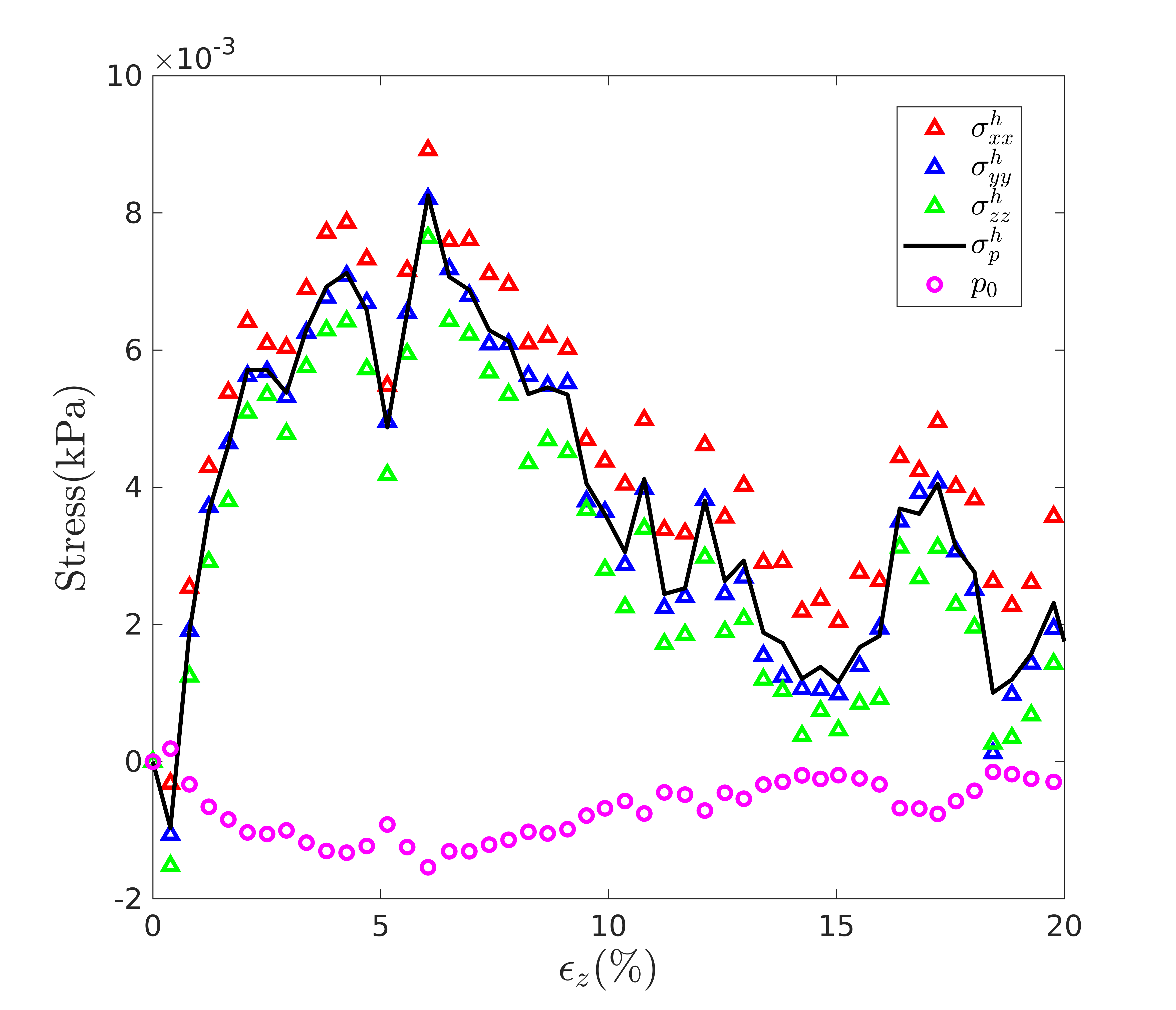}
	\caption{Evolution of the hydrodynamic stress term $\sigma_{ij}^{h}$ and average fluid pressure $p_{0}$ in terms of the axial strain $\epsilon_{z}$ during the shear stage.}
	\label{hydroxyz}
\end{figure}

\begin{figure}
	\centering
	\includegraphics[scale=0.6]{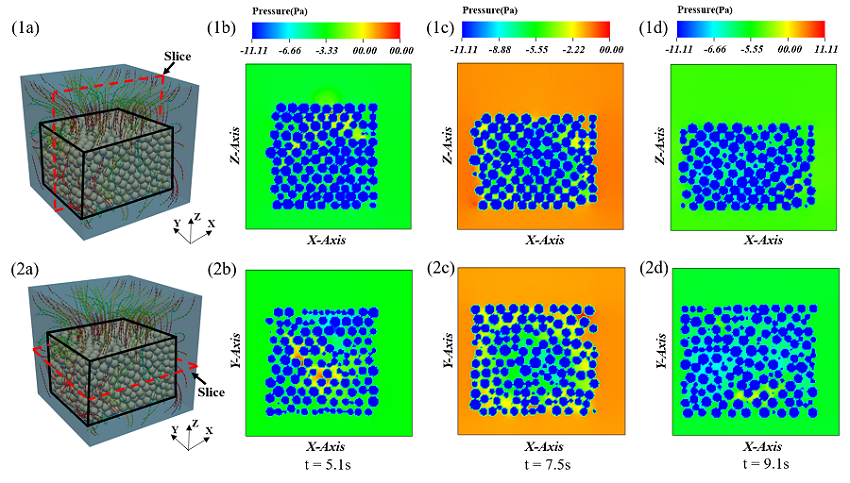}
	\caption{Evolution of the pressure distribution in terms of the simulation time $t$ during the shear stage (time:5-10s). (1a), (2a) shows the position of the slice.}
	\label{drainpredis}
\end{figure}

\begin{figure}
	\centering
	\includegraphics[scale=0.55]{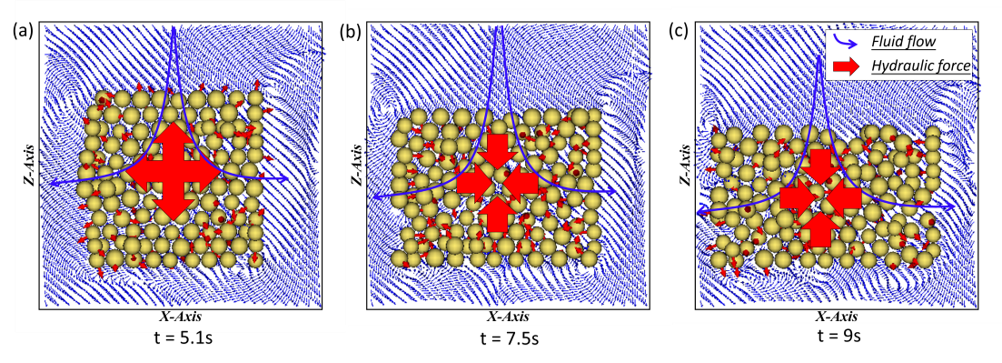}
	\caption{Evolution of the fluid velocity distribution and hydraulic force vector exerting on the particles at different simulation time $t$ during the shear stage (time:5-10s). The red arrow in the center represents the hydraulic effect on the granular assembly. (The arrow pointing outward the RVE represents dilation and pointing inward represents compressing the granular assembly.).}
	\label{drainvector}
\end{figure}

The contribution of the hydrodynamic force term on the effective stress is compared with the average fluid pressure $p_{0}=\sum_{i=1}^{N_{l}}p_{f}^{i}/N_{l}$, where $N_{l}$ is the number of lattice cells inside the RVE and $p_{f}^{i}$ is obtained from Eq.\ref{equa:p}, as shown in Fig. \ref{hydroxyz}. The mean principal stress $\sigma_{p}^{h}$ of the hydrodynamic force term is different from the average fluid pressure: at the beginning of the shear stage, $\sigma_{p}^{h}$ is negative, it increases to the positive value in terms of the axial strain, and then decreases close to zero; however, the average fluid pressure increases from zero, and then decreases to the negative value, at last, close to zero (but less than zero). Firstly, according to the fluid pressure distribution, as shown in Fig. \ref{drainpredis}, the fluid pressure in the pore space is negative at different times. The pressure difference exists between the pore space and the location out of the granular assembly. Thus, the fluid will flow in or out of the granular assembly, which leads the average fluid pressure in the pore space to increase. Secondly, the momentum exchanges are accompanied by fluid flow. Fig. \ref{drainvector} shows the fluid velocity distribution and the hydrodynamic force vector exerting on each particle at a different time: the fluid flow inside the granular assembly from the top and gets out in the lateral direction at a different time; however, the hydrodynamic force subjected to each particle point out of the granular assembly at the beginning of the shear stage(t=5.1s), and then the hydrodynamic force vector point to the center of the granular assembly, which account for the evolution of the hydrodynamic force term $\sigma_{ij}^{h}$ showed in Fig. \ref{hydroxyz}. At last, the fluid pressure in the pore space is close to the pressure outside the granular assembly (see Fig. \ref{drainpredis}). Thus, the fluid flow velocity decreases which lead to the hydrodynamic force term close to zero at the end of the shear stage as shown in Fig. \ref{hydroxyz}. In consequence, the average fluid pressure is much different from the $\sigma_{p}^{h}$. 

\begin{figure}
	\centering
	\includegraphics[scale=0.4]{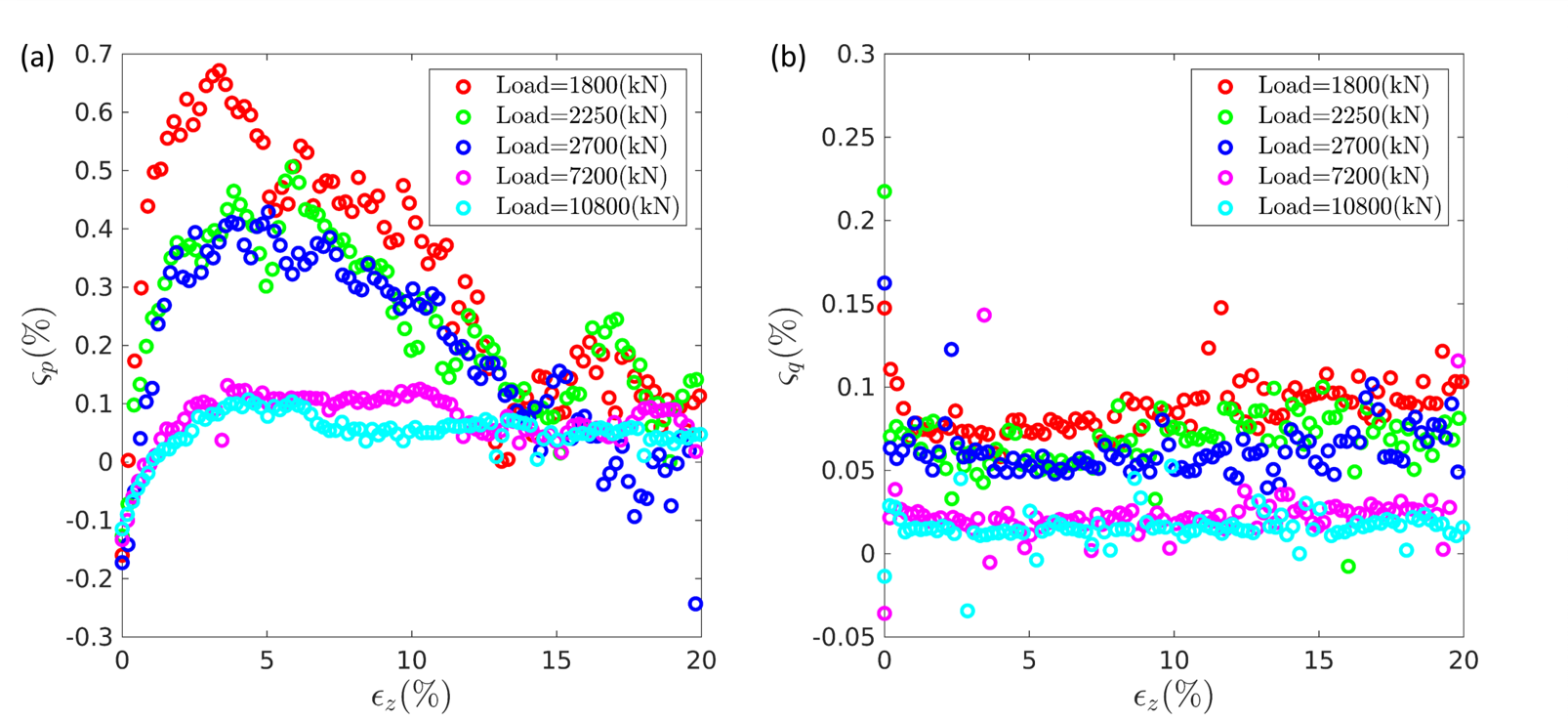}
	\caption{Contribution of $\sigma_{ij}^{h}$ to the mean principal stress and deviatoric stress in terms of the axial strain $\epsilon_{z}$ with different external load: (a) in the x-direction, (b) in the y-direction, (c) in the z-direction. }
	\label{hpqvl}
\end{figure}

The same drained tests were carried out with different plane loads. As shown in Fig. \ref{hpqvl}, the effect of the hydrodynamic force term for the mean principal stress decreases at the beginning of the shear stage and then increases, at last, keeping in a constant value, which is similar to each other. The effect of the hydrodynamic force term on the deviatoric stress decreases with the increase of the external force. Hence, as lower external forces are exerted on the granular assembly, the effect of the hydrodynamic force term is more significant.

\subsubsection{Undrained triaxial test}
\begin{figure}
	\centering
	\includegraphics[scale=0.5]{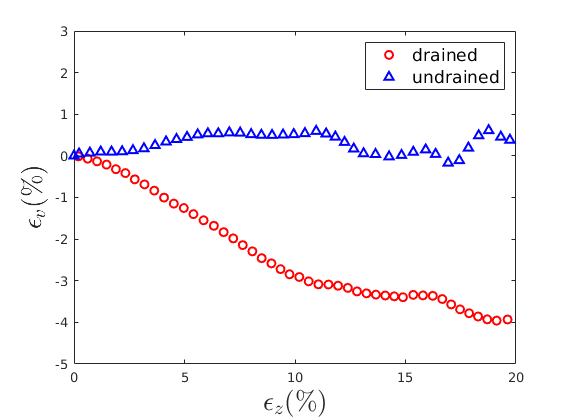}
	\caption{Volume strain $\epsilon_{v}$ in terms of axial strain $\epsilon_{z}$ during the shear stage in drained and undrained tests.}
	\label{svsz}
\end{figure}

The submerged undrained triaxial test is implemented with the same conditions as the drained test, but the fluid cannot pass boundary planes in the shear stage. Due to the high sound speed applied in LBM simulation, the fluid is nearly incompressible, and the volume strain $\epsilon_{v}$ of the granular-fluid mixture barely changes in terms of the axial strain $\epsilon_{z}$ as shown in Fig. \ref{svsz}.  

The compression stage is within 0$-$2 s, and the shear stage is within 2$-$7 s. Except for the contact forces $f_{i}^{S_{k}}$ with the inside particles and the loading force $f_{i}^{l}$, the plane is subjected to the hydrodynamic force $f_{i}^{h,S_{k}}$. Hence, the external force exerted on particles from each plane is given by
\begin{equation}
f_{i}^{S_{k}}=f_{i}^{t,S_{k}}-f_{i}^{l}-f_{i}^{h,S_{k}}.
\end{equation}
The force loading on the fluid-granular mixture domain in each direction is given by
\begin{equation}
f_{i}^{m,S_{k}}=f_{i}^{S_{k}}+f_{i}^{h,S_{k}}.
\end{equation}
Hence, the pressure of the mixture in each direction can be obtained
\begin{equation}
P^{m,S_{k}}=\frac{(f_{i}^{S_{k}}+f_{i}^{h,S_{k}})n_{j}^{S_{k}}}{S_{k}}, 
\end{equation}
the pressure $P_{i}^{e,S_{k}}$ exerting on the granular assembly from each plane can be given using Eq.\ref{equa:wpre}.
The contribution of hydrodynamic force on the mixture domain is from two parts: One is sustained by the fluid given by
\begin{equation}
u_{ij}=\frac{1}{V}f_{i}^{h,S_{k}}x_{j}^{S_{k}},
\end{equation}
where $x_{j}^{S_{k}}$ is the vector from the geometric center of the granular assembly to the center of the plane;
the second is $\sigma_{ij}^{h}$ sustained by the solid part (including in $\langle \sigma_{ij} \rangle$) given by Eq.\ref{equa:hydro}.
The total stress $\sigma_{ij}^{T}$ of the mixture domain consists of both the effective stress subjected to the solid skeleton and the pressure sustained by the fluid,
\begin{equation}
\sigma_{ij}^{T}=\langle\sigma_{ij}\rangle+u_{ij}.
\end{equation}
However, the fluid in the pore space in classical soil mechanics is assumed to be isotropic, and shear stresses are neglected, which is a sound condition when quasi-static conditions are assumed, but not so realistic with highly dynamic granular-water mixture flow. $\delta_{ij}p_{0}$ is assumed to be the contribution of fluid pressure to the mixture domain,
\begin{equation}
p_{0}=\frac{\sum_{c\in RVE}^{}p_{f}^{c}}{N_{in}}-\frac{\sum_{c\in (V-RVE)}^{}p_{f}^{c}}{N_{out}},
\end{equation}
where $c\in RVE$ represents the fluid cell inside the RVE, $c\in (V-RVE)$ represents the fluid cell outside the RVE, $N_{in}$ and $N_{out}$ are the numbers of cells inside or out of the RVE, respectively.

\begin{figure}
	\centering
	\includegraphics[scale=0.33]{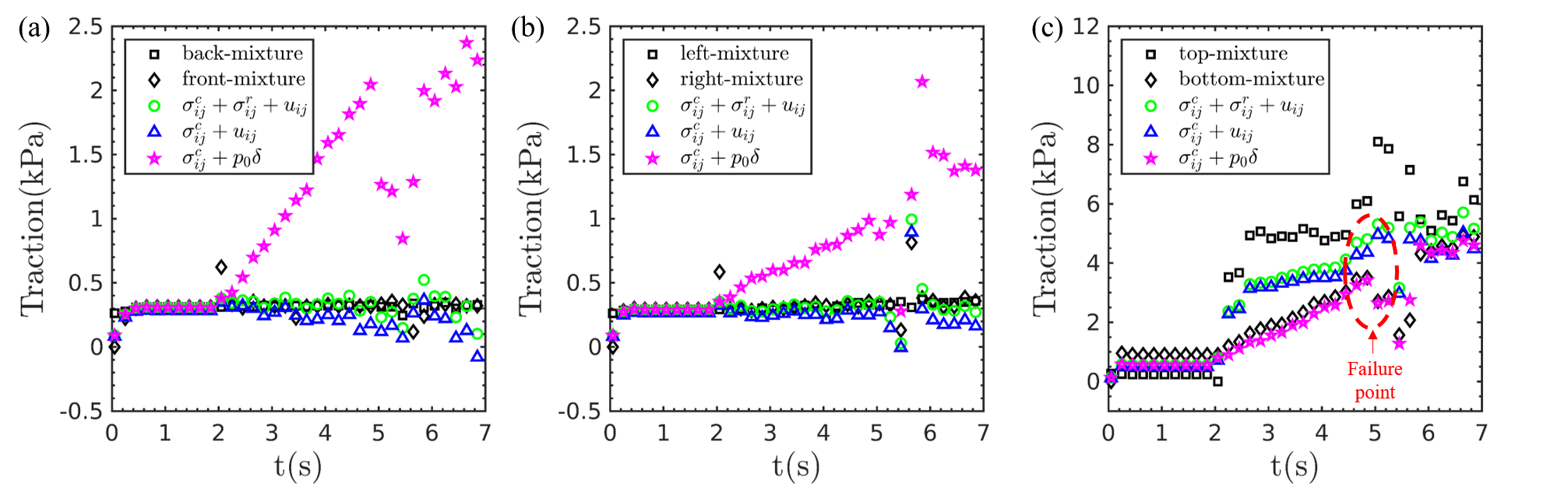}
	\caption{Traction obtained from the averaged stress tensor and pressure of the mixture domain on the boundary during the compression and shear stage (0$<$t$<$10s): (a) in the x-direction, (b) in the y-direction, (c) in the z-direction.}
	\label{stressmixture}
\end{figure}

\begin{figure}
	\centering
	\includegraphics[scale=0.45]{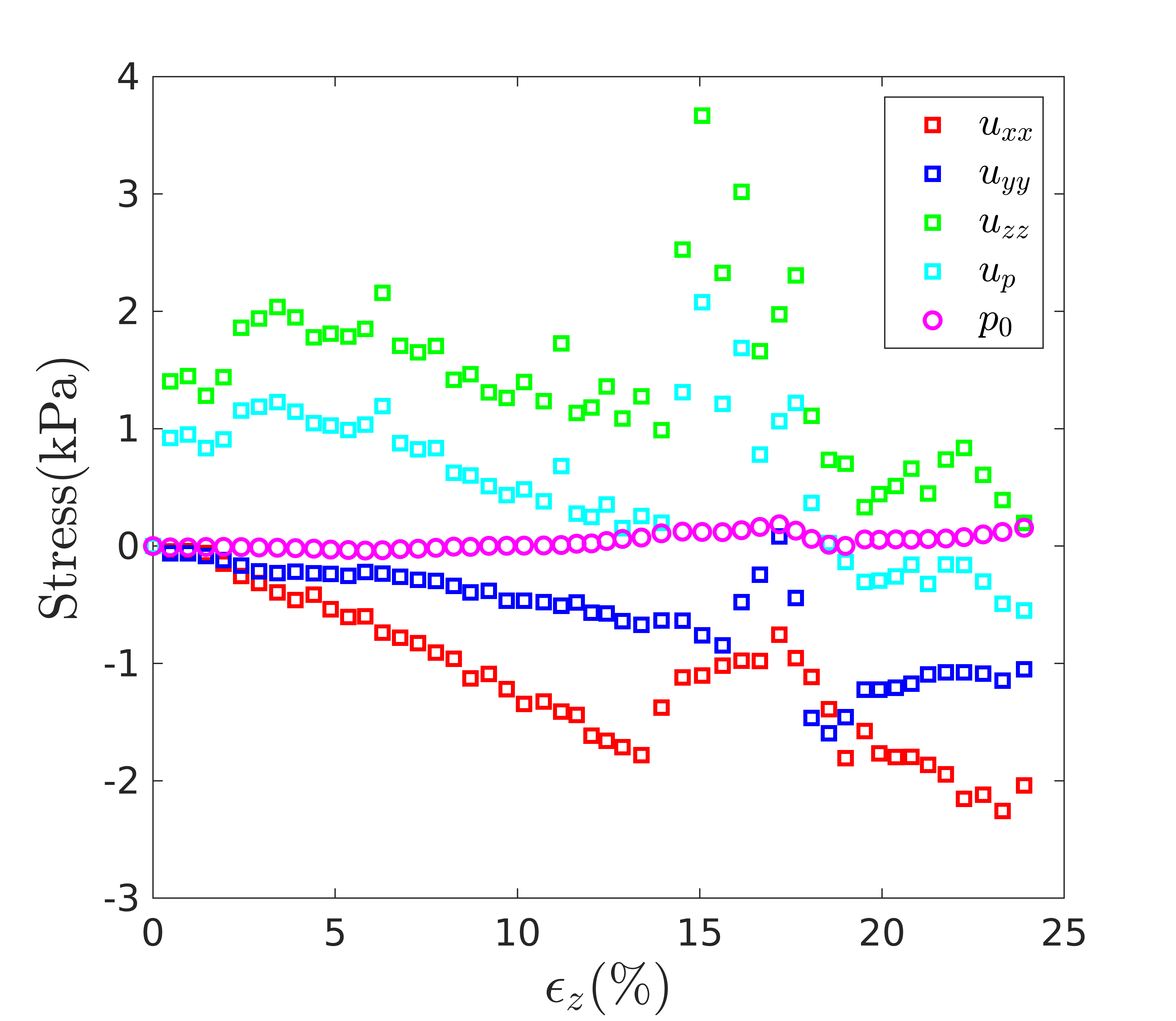}
	\caption{Evolution of the hydrodynamic effect on the mixture domain in terms of $\epsilon_{z}$. $u_{xx}$, $u_{yy}$, $u_{zz}$ represent the three principal stress, $u_{p}$ is the mean principal stress, and $p_{0}$ is the average fluid pressure}
	\label{pu}
\end{figure}

\begin{figure}
	\centering
	\includegraphics[scale=0.64]{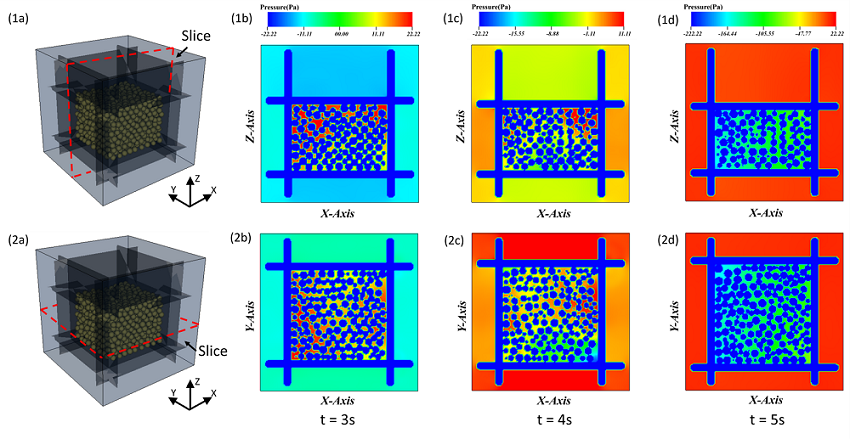}
	\caption{Evolution of the pressure distribution at different simulation time $t$ during the shear stage (time:2-7s). Figs.(1a) and (2a) show the position of the slice.}
	\label{pressuredis}
\end{figure}

As shown in Fig. \ref{stressmixture}, the traction of the total stress of the fluid-granular domain shows good agreement with the average pressure of the mixture domain in each direction. In contrast, the effective stress plus the $p_{0}\delta_{ij}$ shows a significant difference. 
Fig. \ref{pu} shows the evolution of $u_{ij}$ and $p_{0}$ in terms of the axial strain: $u_{ij}$ shows significant anisotropy, in the z-direction, the fluid tends to repose the wall leading the domain to dilate; while in the lateral direction(x, y-direction), the fluid leads the domain compression; the mean principal stress $u_{p}$ indicates that the fluid tends to expand the domain first and then compresses. These phenomena are different from the average fluid pressure $p_{0}$.
Fig. \ref{pressuredis} shows: the fluid pressure distribution is significantly different at each position, which accounts for the anisotropy of the contribution of fluid ($u_{ij}$) on the mixture domain. At the beginning of the shear stage, the internal fluid pressure is larger than the pressure outside the RVE (see Fig. \ref{pressuredis}(2a) and (2b)). In contrast, after 1 s, the fluid pressure in the pore space is close to the pressure outside (out of the granular assembly). At last, the inside fluid pressure is less than the external fluid pressure.  It means that even in the quasi-static regime, the average fluid pressure cannot represent the influence of the fluid on the granular-fluid mixture domain by itself; drag and lubrication effects need to be considered.

\begin{figure}
	\centering
	\includegraphics[scale=0.33]{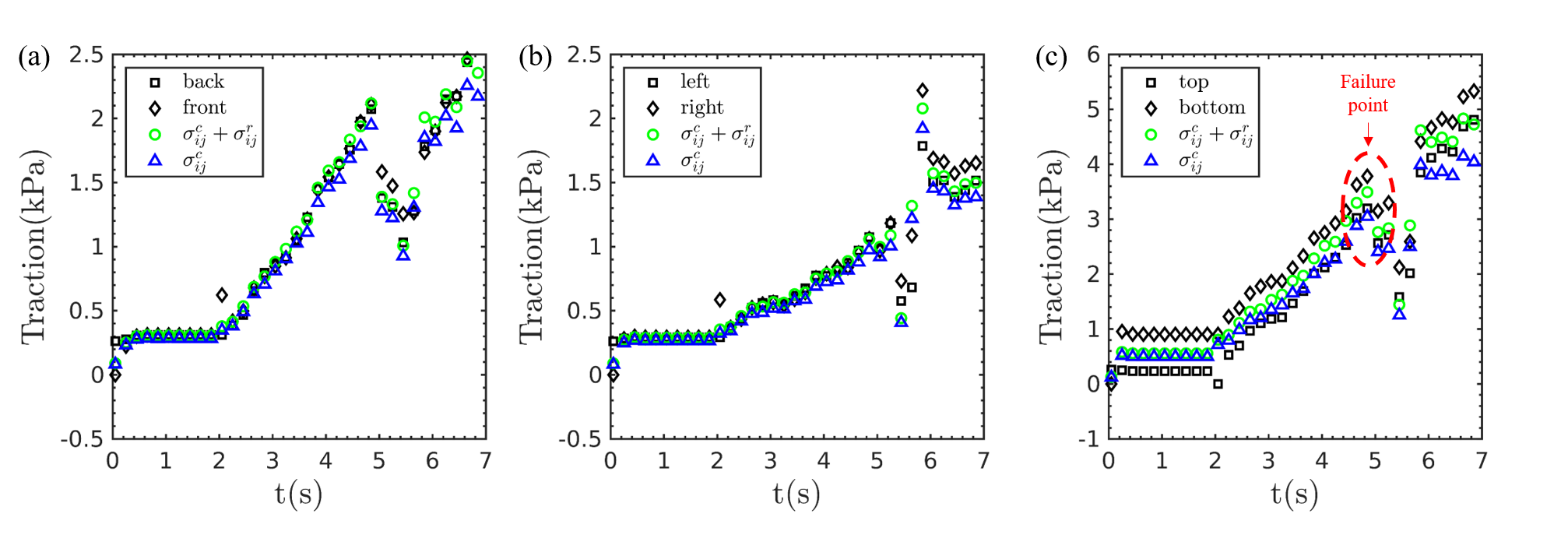}
	\caption{Traction obtained from the averaged stress tensor and pressure of each plane on the compression and shear stage (0$<$t$<$10s): (a) in the x-direction, (b) in the y-direction, (c) in the z-direction.}
	\label{stressundrainl}
\end{figure}

As shown in Fig. \ref{stressundrainl}, the traction of the averaged stress tensor calculated from Eq.\ref{equa:conponents} agrees well with the wall pressure, which is obtained using Eq.\ref{equa:wpre}. The Eq.\ref{equa:conponents} can accurately obtain the effective stress for the undrained submerged triaxial test.

From both the drained and undrained test, we may conclude that the effect of fluid on the granular-fluid assembly consists of two parts: the hydrodynamic stress $u_{ij}$ exerted on the fluid and the hydraulic contribution $\sigma_{ij}^{h}$ to the solid skeleton included in the effective stress $\langle \sigma_{ij} \rangle$; the hydraulic contribution exerted on the solid skeleton influences the solid-phase average stress, and its evolution, while the hydrodynamic stress exerted on the fluid ($u_{ij}$) does not influence the structure of the solid phase. The average fluid pressure in pore space is not equal to the contribution exerted on the solid skeleton or the contribution exerted on the fluid.
\subsection{Submerged granular column collapses}\label{section:SGC}
In this part, we applied the proposed averaged stress tensor and hydrodynamic contribution to the submerged granular column collapse. The physic of submerged granular column collapse is similar to the submerged landslides, hence numerous investigations\cite{pailha2008initiation,rondon2011granular,trulsson2012transition,kumar2017mechanics,lee2019underwater,wang2017dilatancy,yang2020pore,lee2021onset} study this process through both numerical simulations and experiments. An accurate averaged stress tensor in the fluid is crucial to understand the granular flow at macroscopic scale\cite{lacaze2021immersed}. Moreover, the effect of the fluid on the solid deformation could help better understand this natural phenomenon.

\begin{figure}
	\centering
	\includegraphics[scale=0.3]{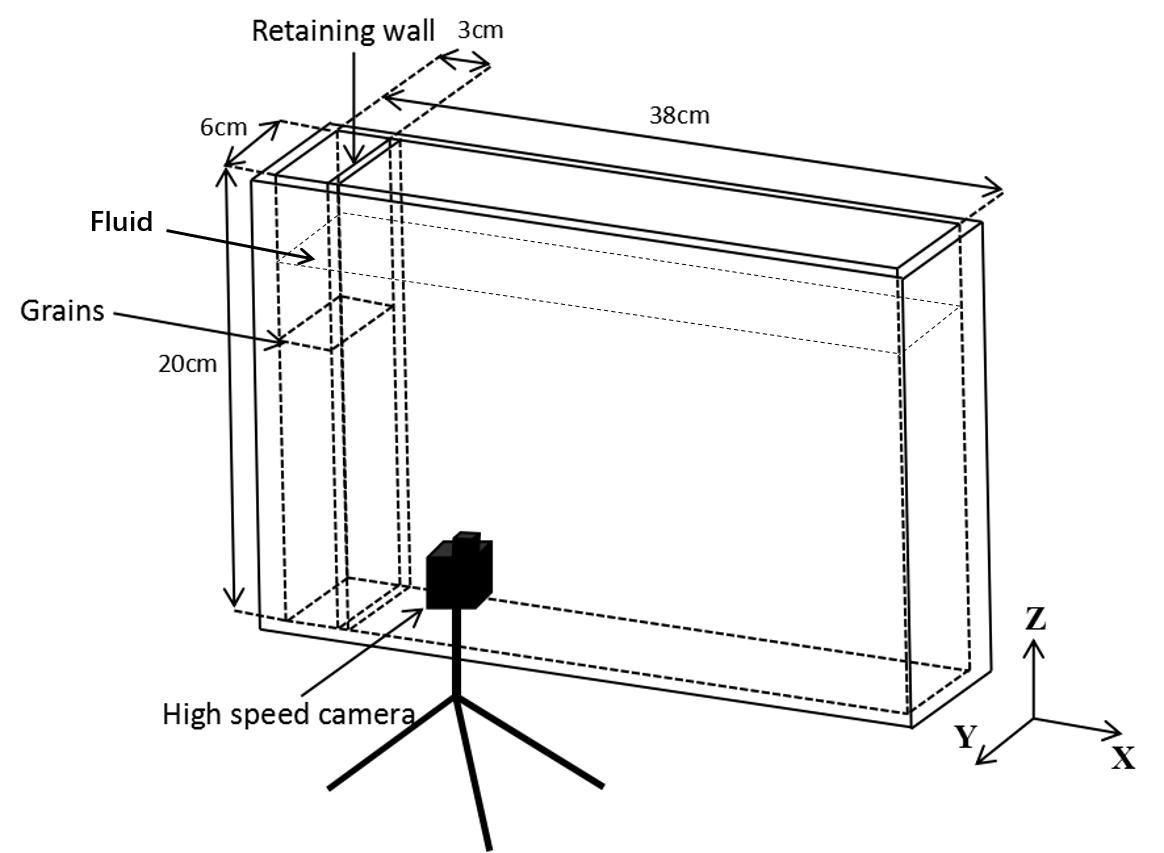}
	\caption{Experimental setup of the submerged granular column collapses.}
	\label{SGCsetup}
\end{figure}

We first perform experiments of submerged granular collapses to validate the LBM-DEM models and then introduce cubic RVEs to obtain macroscopic features, including the averaged stress and hydrodynamic contributions, of the whole domain using the microscopic information. The experimental setup is presented in Fig.\ref{SGCsetup}. 
It is in a 38cm long, 6cm wide, and 20cm high transparent plastic tank. The tank is full of water. Plastic beads are used in both simulation and experiments, the parameters of the plastic beads and water are shown in Table. \ref{SGCpara}. \\
\begin{table}
\centering
\caption{Details of parameters used for the submerged granular collapse. }
\begin{tabular}{c c c}
\hline
parameters&unit&value\\
\hline
Particle density& g/cm$^3$ &1.18\\
Particle radius& cm &0.2-0.25\\
Normal stiffness& g/s$^{2}$ &1$\times$10$^{7}$\\
Tangential stiffness& g/s$^{2}$ &1$\times$10$^{6}$\\
Frictional coefficient&-&0.34\\
Restitution coefficient&-&0.9\\
Dynamic viscosity of fluid& Pa${\cdot}$s &0.001\\
Fluid density& g/cm$^3$ &1\\
\hline
\end{tabular}
\label{SGCpara}
\end{table}

The experimental procedure is as follows. Plastic particles are gently poured into the reservoir delimited by the wall and then pour the liquid into the tank at a level of 16.5 cm. The size of the granular column is 3$\times$6$\times$10.4 cm. When the fluid surface and particle are static, the retaining wall is removed suddenly. A side view will be recorded by a video camera at 30 frames per second.
The LBM-DEM simulation is used to simulate this process, and the LBM grid length is 0.05 cm to make sure that the particle diameter is at least 8 times the grid size. 

\begin{figure}
	\centering
	\includegraphics[scale=0.34]{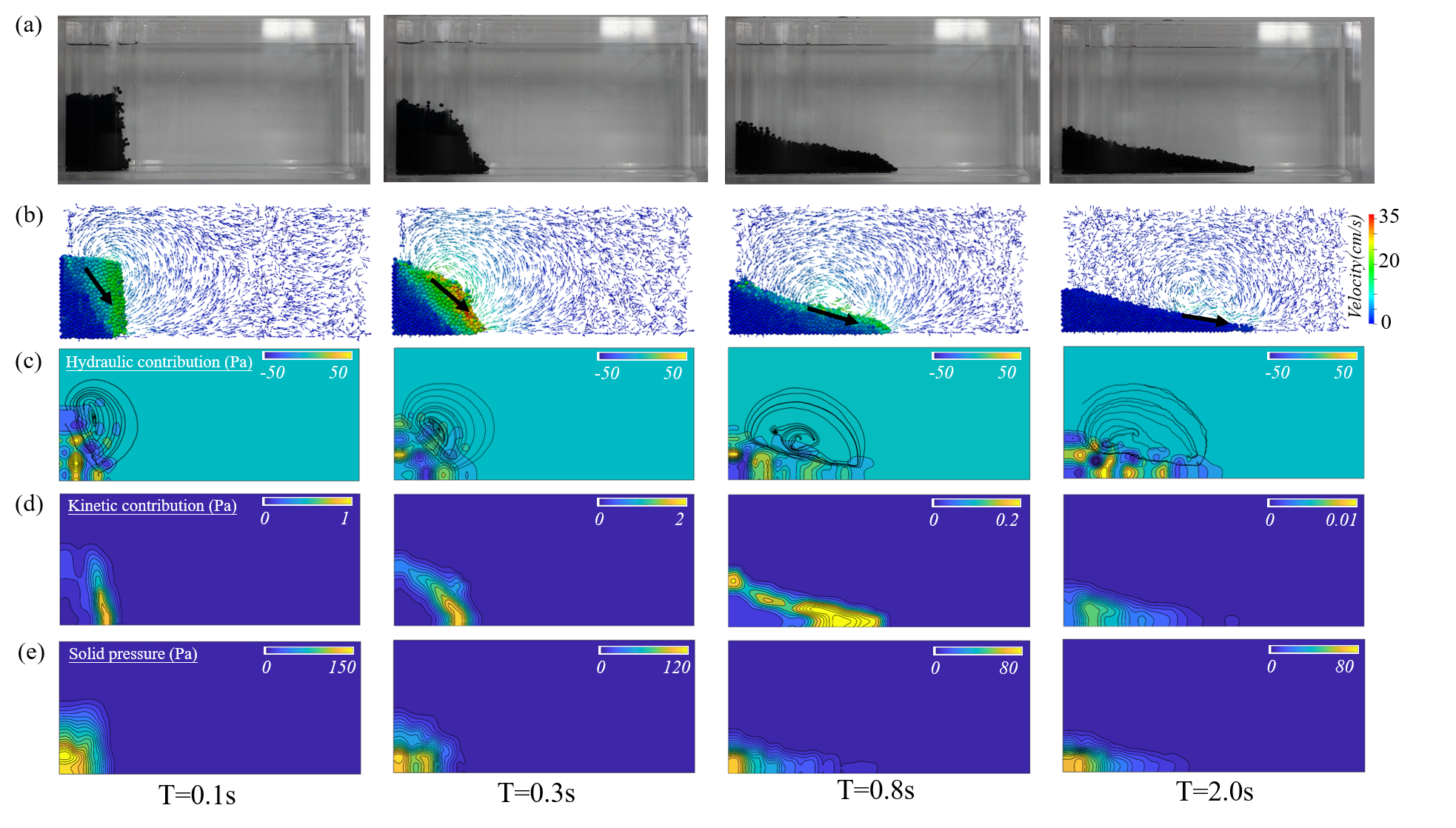}
	\caption{Flow field and particles evolution of submerged granular column collapse at 0.1s,  0.3s, 0.8s and 2.0s: (a) experiment; (b) simulation, the black arrow is the flow direction of the granular; (c) hydrodynamic contribution, $\frac{1}{3}(\sigma_{11}^{h}+\sigma_{22}^{h}+\sigma_{33}^{h})$, on the granular assembly, the black lines is the streamline of fluid;(d)kinetic effect on the averaged stress, $\frac{1}{3}(\sigma_{11}^{k}+\sigma_{22}^{k}+\sigma_{33}^{k})$; (e) total pressure sustained by the solid part. }
	\label{collapserprocess}
\end{figure}

During the granular collapse process, the profile of the granular assembly shows good agreement at different times as shown in Fig.\ref{collapserprocess}(a),(b). Then the RVE is selected with a size of 1.5 cm (around 125 particles). In the work of Yang et al.\cite{yang2020pore}, the dilation and contraction of the granular columns are dependent on the pore fluid pressure, but the real effect of fluid on the granular is the integration of the fluid pressure on the interface between the fluid and solid, at last, this effect is included in the particle-particle interaction in the granular assembly. As shown in Section.\ref{section:SUB}, the contribution of fluid on granular assembly is not equal to the fluid pressure. We use the proposed formula Eq.\ref{equa:hydro} to calculate the hydrodynamic contribution and present it in Fig.\ref{collapserprocess}(c). The effect of fluid on the granular assembly is dependent on the fluid flow, especially the fluid vortex. When the particle is in the front of the flow direction, the fluid tends to compress the solid. However, for particles in the back, fluid tends to play a role to dilate the assembly. Thus, for the same granular column collapse, fluid effects can be completely different in a different location. During the submerged granular collapse, the particle fluctuations are considered as shown in Fig.\ref{collapserprocess}(d), its contribution is calculated by $1/3(\sigma_{11}^{k}+\sigma_{22}^{k}+\sigma_{33}^{k})$. It shows that even in a dynamic system, the kinetic stress is much smaller than the total pressure.

The total pressure of the solid part calculated using Eq.\ref{equa:subs} is shown in Fig.\ref{collapserprocess}(e). This test shows the ability of the proposed method in studying the granular media flow in complex conditions. Combined with the strain field, the stress-strain relation could be investigated in the transient granular flow process. Moreover, this work could help extend the method proposed by Guo et al.\cite{guo2014coupled} to the granular flows in submerged conditions.
\section{Conclusions} \label{sec:conclu}

While the works of \cite{christoffersen1981micromechanical,bagi1996stress,bagi1999microstructural} acquiring the averaged stress tensor, $\langle \sigma_{ij}\rangle=\frac{1}{V}\sum_{N_{c}}f_{i}^{c}l_{j}^{c}$, for the granular assembly is widely used, its accuracy for a granular assembly in various conditions has not been properly validated. Researchers have to increase the number of particles inside the RVE to avoid the boundary-radius-gap term effect. How each force field influences a discrete granular assembly is still only implicitly represented using the internal contact forces (which is the resultant force rather than the source force such as the external and body forces). Hence, it is difficult to distinguish the contributions from different force fields.

This paper derives the expression of the averaged stress tensor for the granular assembly subjected to an arbitrary force field using Hamilton's principle of least action. The contributions of unbalanced local forces, various types of body forces, external forces, contact forces, and kinetic fluctuation are separated and illustrated explicitly. The separation of the contribution of each term can be the bridge between the macroscopic behaviour and the microscopic mechanism with quantitative analysis.  
With the assumption that the boundary radius gap vector is orthogonal to the external surface and its magnitude is equal to the average radius, we propose a method to acquire the boundary-radius-gap term using the grain scale information inside the RVE making the expression of the averaged stress tensor further completed.

Numerical simulations based on DEM and LBM-DEM are performed to reproduce triaxial tests of submerged and dry granular soils to validate the proposed method. The results show that the proposed formula could accurately acquire the averaged stress tensor of discrete assembly subjected to gravity and hydrodynamic forces. The method proposed to obtain the external surface force using the grain scale information also shows good performance.
Investigation of the boundary-radius-gap term provides us following features. The size effect factor and the relative error obey the decaying power-law relationship. As $R_{s}$ increases, the relative error decreases. Hence, to obtain an averaged stress tensor with 95$\%$ accuracy, the size effect factor needs to be larger than 42, and for average stress with 99$\%$ accuracy, the size effect factor needs to be at least 210.

The contribution of fluid to the granular-fluid mixture consists of two parts that exert on both the fluid and the solid. The contribution to the solid part exists in the effective stress tensor, which represents the momentum exchange between the fluid and the solid. The contribution sustained by the fluid part is not the classic pore pressure, which is isotropic, but a stress tensor with different values as the principal components.

Finally, this investigation suggests that the proposed formula is suitable for granular assemblies subjected to arbitrary force fields. With the accurate averaged stress tensor, this work could help better obtain the stress-strain relationship (the constitutive law) of the granular system under complex conditions such as the submerged case or systems subjected to electromagnetic fields, and further, understand the natural phenomenon such as debris flows or produce the new structure of granular materials under control of specific force fields. As the contribution of each force field is procurable, the investigation of the momentum exchange from different phases or sources on the volume scale may provide us with a new perspective to handle multi-scale issues.

A final conclusion drawn from this study is that the concept of effective stress, commonly used in soil mechanics even in flows with rapid deformation, may not always apply. For highly dynamic cases, drag and lubrication effects also play a role and the full hydraulic tensor must be considered. Future research in this area should focus on deriving constitutive forms for this hydraulic stress tensor that can be validated by the proposed LBM-DEM coupling scheme. 
\section*{Declaration of Interests}
The authors report no conflict of interest.

\section*{Acknowledgments}
This work is supported by the National Natural Science Foundation of China (NSFC major project grant NO. 12172305).
We thank Westlake University Supercomputer Center for computational resources and related assistance. The simulations were based on the MECHSYS open source library(\url{http://mechsys.nongnu.org}).


 \bibliographystyle{elsarticle-num} 
 \bibliography{References.bib}





\end{document}